\documentclass[reprint,amsmath,amssymb,aps,prx,nobibnotes,longbibliography]{revtex4-2}
\usepackage{graphicx} \usepackage{dcolumn}
\usepackage{xcolor}
\usepackage{bm}
\usepackage{hyperref}
\usepackage{amsfonts}
\usepackage{amsmath}
\usepackage{physics}
\usepackage{parskip}
\usepackage[normalem]{ulem}

\begin{document}

\title{Physical policy gradient theorem for \textit{in situ} stochastic-adjoint training}

\author{William Tuxbury}
\email{wtuxbury@vt.edu}
\affiliation{Bradley Department of Electrical and Computer Engineering,
Virginia Tech, Blacksburg, Virginia 24061, USA}

\author{Zin Lin}
\affiliation{Bradley Department of Electrical and Computer Engineering,
Virginia Tech, Blacksburg, Virginia 24061, USA}

\begin{abstract}
\textit{In situ} adjoint training extracts parameter gradients directly from measurement, but has so far been limited to reciprocal or restricted systems. We introduce the physical counterpart of the policy gradient theorem: a stochastic-adjoint gradient estimator that lifts these constraints by trading reciprocity for nondegenerate diffusion. As validation, we train a nonlinear resonator network, whose own dynamics supply the policy, against antagonistic temporal modulations with gradients from measured stochastic trajectories alone, without finite differences or a separate adjoint experiment.
\end{abstract}

\maketitle

\section{Introduction}\label{sec:intro}
Physical neural networks propagate and process information according to the physics of their substrate. Complexity from multi-path interference, nonlinearity, and parametric encoding has been harnessed to train physical systems that classify data and realize target functionalities~\cite{psaltis1990holography, shen2017deep, lin2018all, Wetzstein2020DeepOptics, Shastri2021PhotonicsAI, DMW2025, wright2022, xue2024fully, Fei2024, Fleury2025,GWLK2025}. Conventionally, these devices are simulated and optimized \textit{in silico} prior to deployment. This paradigm, however, introduces a mismatch between model and device, which grows more severe as the substrate incorporates the very complexity that enhances its expressivity: features such as nonlinearity, time modulation, and stochasticity.

A parallel lineage executes gradient-based training \textit{in situ} on reconfigurable physical platforms, reducing reliance on a complete digital model and thereby minimizing the sim-to-real gap. The operative principle of this approach is reciprocity: excitation and observation sites may be interchanged without altering the system's intrinsic dynamics, enabling the same hardware to emulate the adjoint of its own equations of motion~\cite{bendsoe2004topology, molesky2018inverse,pai2023,GWLK2025}. Adjoint methods are especially efficient because they recover the full gradient with respect to $\mathcal{P}$ parameters using only one forward and one adjoint measurement, rather than the $\mathcal{P}+1$ measurements required by forward finite differences. Although \textit{in situ} adjoint protocols have been demonstrated across a broad class of reciprocal media~\cite{hughes2018,pai2023,GWLK2025,kwon2026insitutimedomainphysicaladjoint} and extended to specific nonlinear defects and symmetries~\cite{guillamon2026insituadjointwavecontrol,DMW2025,bosch2026unifyingphysicalbackpropagation,li2026insituadjointprotocolsnonlinear}, they remain inapplicable to unrestricted dynamics, such as arbitrary nonlinear, time-modulated, and stochastic systems.

To overcome these limitations, we approach \textit{in situ} adjoint training with a complementary formulation. Our approach is inspired by reinforcement learning in uncertain environments, wherein an agent interacting with the environment learns a policy that maximizes the average accumulated reward, or total expected return. Prevailing methods parameterize the policy and optimize those parameters using policy gradient methods. The workhorse behind breakthrough developments in reinforcement learning, from the foundational REINFORCE algorithm~\cite{williams1992simple} to modern methods including trust region policy optimization (TRPO)~\cite{schulman2015trust} and proximal policy optimization (PPO)~\cite{schulman2017proximalpolicyoptimizationalgorithms}, is the policy gradient theorem (PGT)~\cite{NIPS1999_464d828b}. In the PGT, the gradient of the expected return is expressed as the expectation value of the return-to-go weighted by the score of the policy. Crucially for our purposes, the PGT differentiates only the policy with respect to parameters along sampled trajectories. The alternative, deterministic backpropagation, would require emulating the transpose Jacobian of the state dynamics, which cannot, in general, be implemented through the same forward physical processes~\cite{guillamon2026insituadjointwavecontrol, bosch2026unifyingphysicalbackpropagation}.

This perspective raises three questions: can the intrinsic dynamics of an arbitrary nonlinear stochastic substrate embody a policy? Does an analog of the PGT exist for such a physically embodied policy? And importantly, can its gradient be evaluated \textit{in situ}? In this work, we answer these questions by introducing a score-based policy gradient theorem for continuous-time stochastic physical dynamics in direct analogy with PGT, enabling \textit{in situ} adjoint training of physically embodied policies.

Continuous-time reinforcement learning has been developed for deterministic dynamics~\cite{doya2000reinforcement,munos2006policy} and for controlled diffusions~\cite{wang2020reinforcement,jia2022policy}, but the policy in each case is a parameterized construct evaluated over the dynamics. Here, the dynamics are the policy. Our formulation rests on three ingredients: the Fokker--Planck equation~\cite{risken1989fokker}, the Feynman--Kac theorem~\cite{shreve1991brownian}, and It\^o's lemma~\cite{ito1951stochastic}. Although individual trajectories of a stochastic process are random, their distribution evolves deterministically according to the Fokker--Planck equation. Specifying a performance objective as an expected return generates a corresponding adjoint equation whose solution determines the parameter gradient. Then, the Feynman--Kac theorem identifies the adjoint solution with the expected return-to-go, which can be estimated from the forward trajectories without additional adjoint experiments. Backward stochastic differential equation methods, by comparison, represent the adjoint pathwise and require a separate numerical solve~\cite{pardoux1990adapted,exarchos2018stochastic}. Lastly, taking expectations of It\^o's lemma connects the generator of the It\^o process with the rate of change of an observable. When this observable is chosen as the adjoint field (return-to-go), parameter sensitivity is recast via the score function associated with the transition kernel, which itself is the physically embodied policy.

Unlike previous \textit{in situ} adjoint protocols~\cite{pai2023,GWLK2025,DMW2025,guillamon2026insituadjointwavecontrol,li2026insituadjointprotocolsnonlinear}, our formulation applies to arbitrary It\^o processes by trading reciprocity for nondegenerate diffusion, intrinsic to any physical device. As a validation, we simulate the training of a physically embodied policy in a driven resonator network with Kerr nonlinearity and time-modulated coupling, modeled with temporal coupled mode theory (TCMT)~\cite{joannopoulos2008photonic,zhu2013temporal}, extended to include Langevin noise, detailed in Appendix~\ref{app:tcmt} and~\cite{SM}. More broadly, our framework recasts a physical system’s intrinsic dynamics as a trainable policy, opening a route toward online reinforcement learning in matter.

\section{Framework}\label{sec:framework}
Consider the generic It\^o process,
\begin{equation}\label{eq:main_ito}
dX_t=\mu\left(X_t,t\right)\,dt+\sigma\left(X_t,t\right)\,dW_t,
\end{equation}
with $X_t\in\mathcal{X}\subseteq\mathbb{R}^\mathcal{M}$, $\mu(x,t):\mathcal{X}\times[t_0,T]\rightarrow\mathbb{R}^\mathcal{M}$, $\sigma(x,t):\mathcal{X}\times[t_0,T]\rightarrow\mathbb{R}^{\mathcal{M}\times\mathcal{W}}$, and $W_t$ a standard $\mathcal{W}$-dimensional Wiener process. The drift $\mu(X_t,t;\theta)=\mu(X_t,t)$ and diffusion $Q(X_t,t;\theta)=\sigma(X_t,t)\sigma^\top(X_t,t)/2$ also depend on the parameter vector $\theta\in\mathbb{R}^\mathcal{P}$. We define an objective function, $J(\theta)=\mathbb{E}[g_T(X_T)+\int_{t_0}^Tdt\,g(X_t,t)]$, as an expectation value involving both terminal $g_T(X_T)$ and continuing $g(X_t,t)$ rewards. The goal is to efficiently compute the gradient $\nabla_\theta J$ by sampling the original stochastic differential equation over $\mathcal{E}$ trajectories. The distribution of the process solves the corresponding Fokker--Planck equation, which we write in its generator form, $\partial_tp=\mathcal{L}_t^\dagger p$ with $p(x,t_0)=p_0(x;\theta)$.
Neglecting initial sensitivity and explicit parameter dependency in the rewards (see Appendix~\ref{app:neglected}), a reduced form of the standard adjoint-method result expresses the gradient as
\begin{equation}\label{eq:reduced_gradient}
\nabla_\theta J=\mathbb{E}\left[\int_{t_0}^Tdt\,\left(\partial_\theta\mathcal{L}_t\lambda\right)\left(X_t,t\right)\right],
\end{equation}
where $\lambda(x,t)$ solves the adjoint partial differential equation, $\partial_t\lambda+\mathcal{L}_t\lambda+g=0$ with terminal condition $\lambda(x,T)=g_T(x)$~\cite{li2004adjoint,thuburn2005climate,kaltenbacher2018parameter}. A simplified form of the well-known Feynman--Kac theorem~\cite{shreve1991brownian} allows the adjoint variable to be sampled from the same It\^o process as that under stochastic optimization,
$\lambda(x,t)=\mathbb{E}[g_T(X_T)+\int_t^Td\tau\,g(X_\tau,\tau)\big\vert X_t=x]$. Since $\lambda$ is the conditional expected return-to-go, it can be evaluated at $\mathcal{T}+1$ discrete time steps $t_n\equiv t_0+n\Delta t$ ($n=0,\dots,\mathcal{T}$), with $t_\mathcal{T}=T$, via the forward trajectories as the unbiased return-to-go,
\begin{equation}\label{eq:adjoint_estimator}
\hat\lambda_n\equiv g_T\left(X_{t_\mathcal{T}}\right)+\sum_{k=n}^{\mathcal{T}-1}g(X_{t_k}, t_k)\Delta t.
\end{equation}
However, $\mathcal{L}_t$ is a differential operator requiring state-space derivatives of the adjoint field, which pointwise estimates do not supply. The same obstacle arises in simulation, where it is met by estimating those derivatives through pathwise differentiation of the trajectories~\cite{wang2024efficient}; that route is unavailable \textit{in situ}, as it requires co-simulating sensitivity equations alongside the state. Therefore, we aim to rewrite Eq.~\eqref{eq:reduced_gradient} in a form amenable to sampling $\hat{\lambda}_n$ from the forward trajectories alone.

For a Markov process, state transitions over a step $\Delta t$ are set by the kernel $p_\theta^{\Delta t}(x'\vert x)$. Taking expectations of It\^o's lemma applied to an arbitrary time-dependent function $\varphi(x,t)$ (Dynkin's formula~\cite{dynkin2012theory}), the rate of change of its conditional expectation along the process is $(\partial_t\varphi+\mathcal{L}_t\varphi)\!(x,t)=\lim_{\scriptscriptstyle\Delta t\to0}\frac{1}{\Delta t}[\int dx'\,\varphi(x',t+\Delta t)p_\theta^{\Delta t}(x'\vert x)-\varphi(x,t)]$. Differentiating in $\theta$, using the log-trick, and taking $\varphi=\lambda$, we have
\begin{align}\label{eq:Markov_derivative}
\left(\partial_\theta\mathcal{L}_t\lambda\right)&(x,t)=\\
\lim_{\scriptscriptstyle\Delta t\to0}&\tfrac{1}{\Delta t}\mathbb{E}\left[\lambda\left(X_{t+\Delta t},t+\Delta t\right)U(x,X_{t+\Delta t};\theta)\!\mid\! X_t=x\right],\notag
\end{align}
where we have introduced the score function, $U(x,x';\theta)\equiv\partial_\theta\ln p_\theta^{\Delta t}(x'\vert x)$, hereafter suppressing its $\theta$ argument along sampled trajectories. Finally, from Eqs.~\eqref{eq:reduced_gradient} and \eqref{eq:Markov_derivative}, we obtain the physical policy gradient theorem (PPGT),
\begin{align}\label{eq:PPGT_gradient}
\nabla_\theta J&=\mathbb{E}\left[\mathcal{G}\left(\{X_{t_n}\}_{n=0}^\mathcal{T}\right)\right],\\
\text{where}\quad\mathcal{G}&\left(\{X_{t_n}\}_{n=0}^\mathcal{T}\right)=\sum_{n=0}^{\mathcal{T}-1}\hat\lambda_{n+1}U(X_{t_n},X_{t_{n+1}}).\notag
\end{align}
The adjoint variable has been replaced with its forward-sampled estimate from Eq.~\eqref{eq:adjoint_estimator}. Equation~\eqref{eq:PPGT_gradient}, consistent in the limit $\Delta t\rightarrow0$, is the physically embodied counterpart of the policy gradient theorem~\cite{NIPS1999_464d828b}, wherein the gradient estimator can be expressed $\nabla_\theta J = \mathbb{E}\big[\textstyle\sum_k G_k\,\partial_\theta\ln\pi_\theta(A_k\vert S_k)\big]$. The transition kernel $p_\theta^{\Delta t}(x'\vert x)$ plays the role of the policy $\pi_\theta$, its score $U$ is $\partial_\theta\ln\pi_\theta$, the sampled Feynman--Kac adjoint $\hat\lambda$ is the return-to-go $G_k$, and $X_t$ is the continuous-time analog of $S_k$. The physical policy is no longer an abstracted rule set, but is instead baked into the dynamics themselves. The substrate's own stochastic evolution supplies the action distribution, and the state transition, $X_t\rightarrow X_{t+\Delta t}$, is the action, $A_k$. When $\theta$ only enters the drift $\mu$, the score function of an It\^o process following Euler--Maruyama updates (discretized form of Eq.~\eqref{eq:main_ito}~\cite{SM}; $dt\rightarrow\Delta t$) has the closed form $U(x, x';\theta)=\tfrac{1}{2}r(x,x')^\top Q^{-1}\partial_\theta\mu(x,t)$, with the residual $r(x,x')\equiv x'-x-\mu(x,t)\Delta t$. Nondegenerate diffusion must span each parameter perturbation of the dynamics over a kernel transition, since otherwise the sensitivity leaves no trace in the score. See Appendices~\ref{app:score} and~\ref{app:complete} for the complete forms of the score function, including diffusion $Q$, and the PPGT, with all terms reincorporated.

\begin{figure}[t]
\centering
\includegraphics[width=\columnwidth]{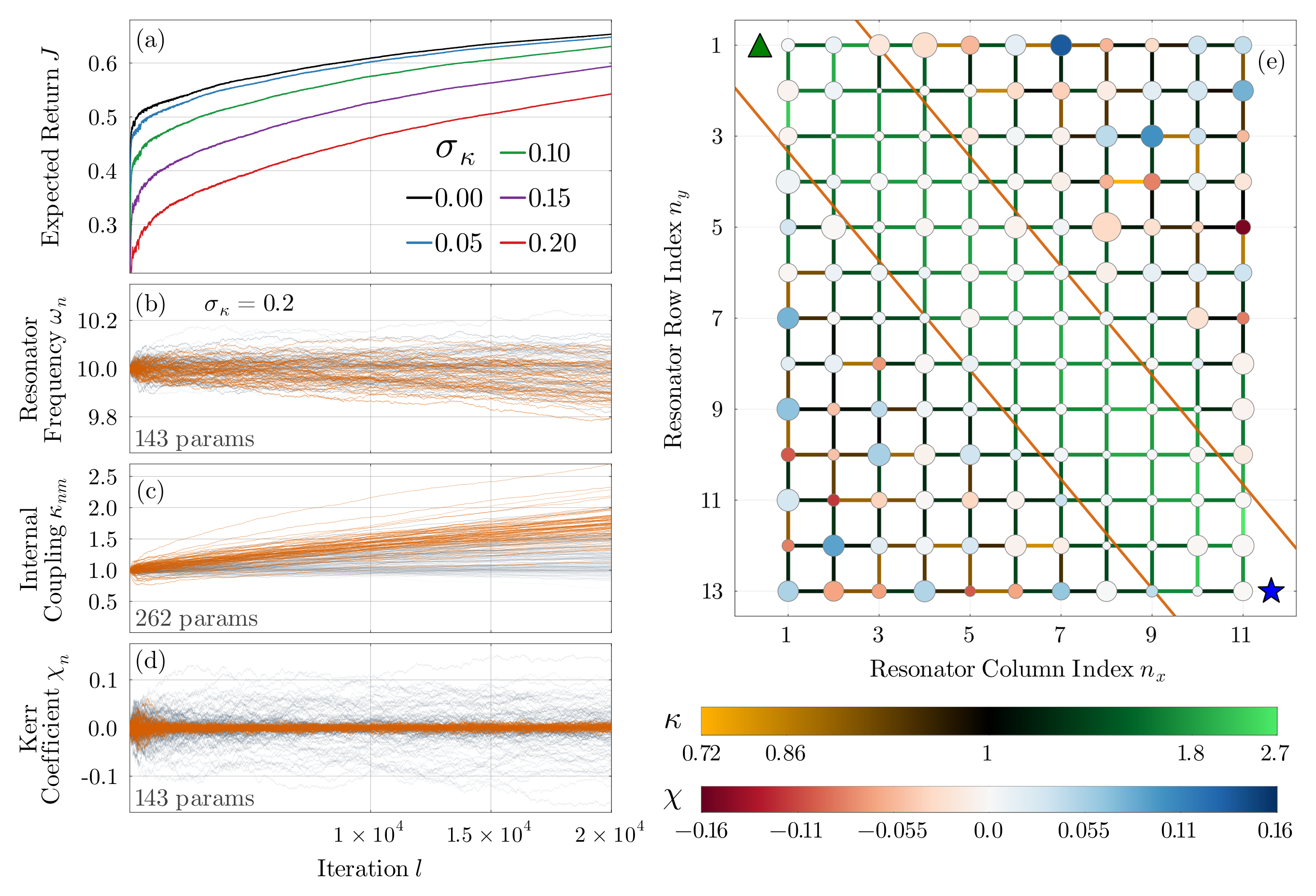}
\caption{\label{fig:convergence}
  \textbf{Physical Policy Training}
  \textbf{(a)}~Expected return versus training iteration $l$ for $\sigma_\kappa$ values indicated by the legend. \textbf{(b)}~Resonator frequency, \textbf{(c)}~internal coupling, and \textbf{(d)}~Kerr coefficient parameters versus training iteration $l$ for $\sigma_\kappa=0.2$. Curve opacities are set by time-integrated field exposures (see Appendix~\ref{app:opt}) with vermilion hue if they lie at least half inside the corridor in \textbf{(e)}, and slate otherwise. \textbf{(e)}~Network diagram of the trained optimum for $\sigma_\kappa=0.2$ with port-to-port boundaries enclosing a corridor three sites wide, indicating an effective waveguide, where the green triangle (blue star) marks the input (output) ports. Disk radii increase with resonator frequency detuning; bond color encodes the internal coupling magnitude, compressed on asymmetric square-root scales about their initial value $\kappa_{nm}=1$; disk color represents the Kerr coefficient. Color bars are shown beneath.}
\end{figure}

\section{Results}\label{sec:results}
As validation, we deploy PPGT in an $11\times13$ rectangular coupled-resonator lattice, consisting of $N=143$ sites, modeled using the TCMT effective Hamiltonian, $H_\mathrm{eff}$, with nearest-neighbor internal coupling and two leads attached at opposite corners, whose orthonormal channel basis elements are denoted $\ket{e_{1}}$ and $\ket{e_{2}}$. All coupling strengths are taken real and symmetric, with reciprocity strictly violated by explicit state- and time-dependent dynamics due to Kerr nonlinearity and modulated internal coupling. Hence, this setting is inaccessible to reciprocity-based methods. The external coupling rates are symmetric and normalized, $\gamma_e=1$, while radiative decay rates are assumed uniform, $\gamma=0.02$. The $143$ resonator frequencies $\omega_n(\theta)$, $262$ static internal coupling strengths $\kappa_{nm}(\theta)$, and $143$ Kerr coefficients $\chi_n(\theta)$ are all trainable and real for a total of $\mathcal{P}=548$ parameters. Here, they enter only through the drift, so the score is readily obtained from $\partial_\theta H_\mathrm{eff}$ (see Appendix~\ref{app:tcmt}).

The system is driven by a harmonic source at the first lead, $\ket{s^+}=10\,e^{-i\omega t}\ket{e_1}$ for $t\ge t_0=0$, from zeroed initial condition $p_0(x)=\delta^{\mathcal{M}}(x)$, with $\mathcal{M}=2N$. The return is taken as the transmitted power to the second lead over $t_0\le t\le T$, with $\omega=10$ and $T=20$, so the continuing reward is $g(x,t)=\lvert\braket{e_2}{s^-(t)}/\braket{e_1}{s^+(t)}\rvert^2$ and zero terminal reward $g_T(x)=0$, where $\ket{s^-(t)}$ denotes the outgoing wave. During training, additive Wiener processes were incorporated into all open channels with identical amplitude $\sigma_W=0.5$, without which the diffusion is singular and the score diverges, $U\sim Q^{-1}$. We also considered a class of antagonistic, temporally-correlated, internal coupling modulations $\delta\kappa_{nm}(t)$, drawn independently per bond and trajectory (see Appendix~\ref{app:tcmt}).

Several lattices were trained under various antagonistic modulation amplitudes $\sigma_\kappa$ ranging from $0$ to $0.2$. The improvement in expected returns is plotted in Fig.~\ref{fig:convergence}(a) versus iteration $l\leq20{,}000$. Each iteration was supplied with a trajectory budget $1{,}000\le\mathcal{E}\le30{,}000$ (see Appendix~\ref{app:opt} for optimization details). While gradual progress persisted for tens of thousands of iterations, most improvement happened early and with a relatively cheap trajectory budget. For instance, in the $\sigma_\kappa=0$ optimization $40\%$ of the eventual gain in expected return was achieved within the first $15$ iterations and an additional $20\%$ within the next $60$ before the budget was increased from $\mathcal{E}=1{,}000$. Parameter trajectories for $\sigma_\kappa=0.2$ are shown in Fig.~\ref{fig:convergence}(b--d), visualized at $l=20{,}000$ in Fig.~\ref{fig:convergence}(e), by which point progress had not yet plateaued. The latter shows a band of enhanced coupling along the diagonal (green bonds) with detuned frequencies on the flanks forming an effective waveguide. Little use was made of the available Kerr coefficients (high-exposure sites have $\chi_n\approx0$ in Fig.~\ref{fig:convergence}(d)), which reflects the choice of performance objective. Transmission alone does not require nonlinearity, unlike fundamentally nonreciprocal tasks, such as isolation.

\begin{figure}[t]
\centering
\includegraphics[width=\columnwidth]{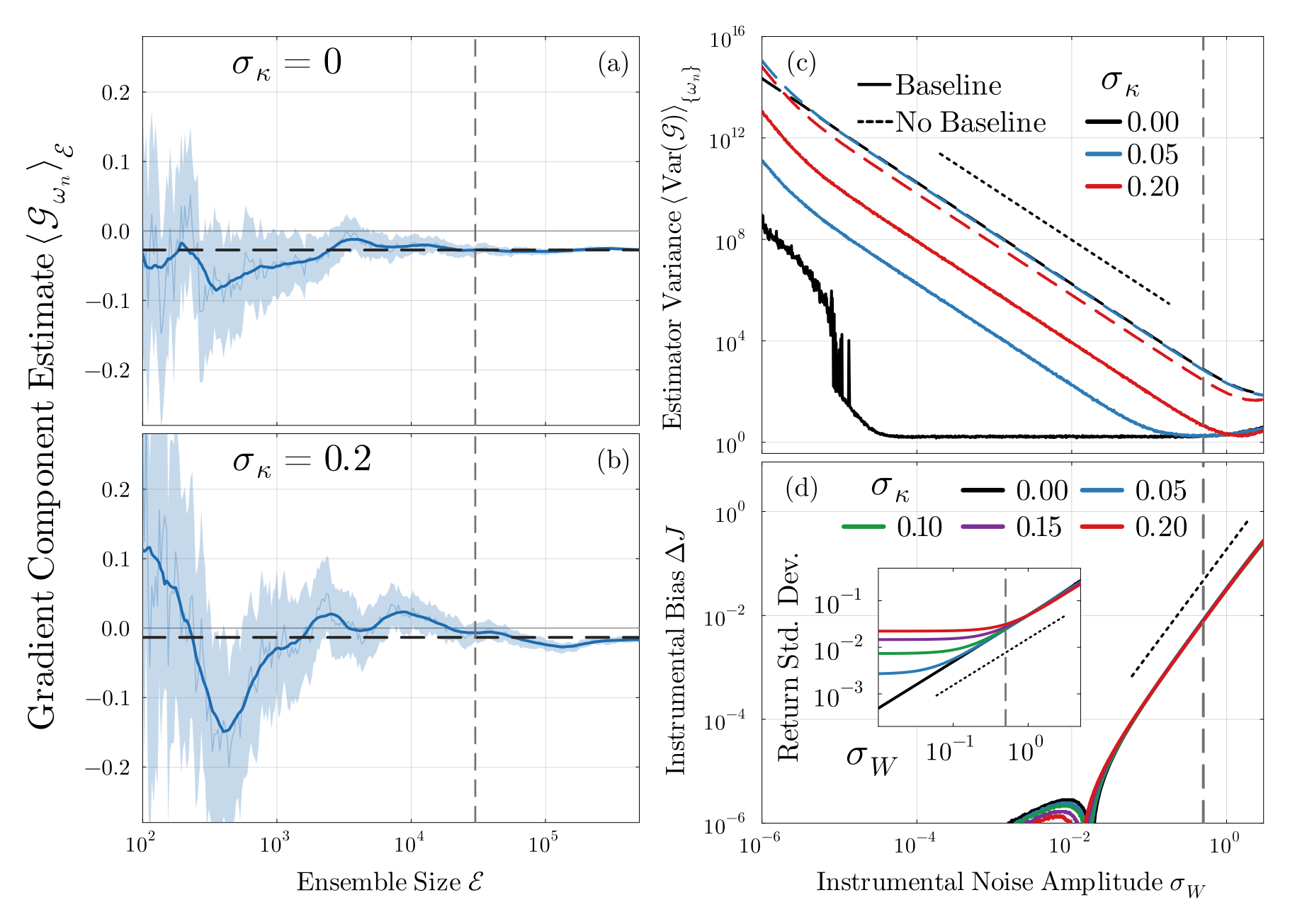}
\caption{\label{fig:estimator}
  \textbf{Estimator Properties} Convergence of single components of the gradient estimate associated with resonator frequency for \textbf{(a)}~$\sigma_\kappa=0$ and \textbf{(b)}~$\sigma_\kappa=0.2$ versus the number of trajectories accumulated in the ensemble average. The horizontal dashed lines indicate the finite-difference values. The solid blue lines are moving averages of the running estimate, shown unsmoothed and faded underneath, while the light-blue shaded band indicates its standard error. \textbf{(c)}~Gradient estimate variance versus instrumental noise amplitude with (solid) and without (dashed) baselines, shown for $\sigma_\kappa=0,0.05,$ and $0.2$ and with the dotted trend line $\sim\sigma_W^{-2}$. \textbf{(d)}~Bias in the expected return at the final iteration due to $\sigma_W\ne0$ for each $\sigma_\kappa$ in the legend and with the dotted trend line $\sim\sigma_W^{2}$. Inset: Standard deviation of the return with dotted trend line $\sim\sigma_W$, showing a crossover as instrumental noise overtakes antagonistic noise as the dominant source of return variance. The vertical dashed lines in \textbf{(a)}, \textbf{(b)} and \textbf{(c)}, \textbf{(d)} respectively indicate the maximum trajectory budget and instrumental noise amplitude $\sigma_W$ in the optimizations.}
\end{figure}

Estimates of $\partial_{\omega_n}\!J$ for the $\sigma_\kappa=0$ and $0.2$ optimizations are plotted in Fig.~\ref{fig:estimator}(a,b), at respective mid-training iterations and site indices selected to favor the resolvability of systematic bias, avoiding pathologies (e.g., lead-coupled sites or detuned flanks). The estimates converge to the finite-difference values, computed using $\mathcal{E}=1{,}000$ with common noise realizations across the differenced legs. Notably, over $\mathcal{P}=548$ dimensions a noisy update is still informative enough for progress; per-component accuracy is not required. Moreover, noisy gradients are often advantageous in stochastic optimization~\cite{bottou2010large}.

The variance scaling of the PPGT estimator is analyzed in Fig.~\ref{fig:estimator}(c). The estimator variance, computed at a representative mid-training iteration using $\mathcal{E}=1{,}000$ and averaged over all resonator frequencies, denoted $\langle\,\cdot\,\rangle_{\{\omega_n\}}$, is plotted versus Wiener noise amplitude $\sigma_W$ for various $\sigma_\kappa$. From Eq.~\eqref{eq:main_ito}, $Q^{-1}\sim\sigma_W^{-2}$ and $r\sim\sigma_W$, so $U\propto r^\top Q^{-1}\sim\sigma_W^{-1}$, then assuming $\lambda$ scales constant with $\sigma_W$, from Eq.~\eqref{eq:PPGT_gradient} we predict the observed scaling, $\mathrm{Var}(\mathcal{G})\sim (\sum \lambda \,U)^2\sim \sigma_W^{-2}$ (dotted trend line). The baseline reduces estimator variance by several orders of magnitude, to the extent that in the best case of $\sigma_\kappa=0$ it is suppressed to a constant. In all scenarios, $\lim_{\sigma_W\rightarrow0}\mathrm{Var}(\mathcal{G})\rightarrow\infty$ as $Q$ becomes ill-conditioned.

Nondegenerate diffusion is a requirement of PPGT, and in fact instrumenting noise can suppress estimator variance. However, the same noise has the opposite effect on the measured return. This is quantified by the instrumental bias $\Delta J\equiv\lvert J(\sigma_W)-J(0)\rvert$, computed using $\mathcal{E}=10{,}000$ as paired differences against $\sigma_W=0$ under common noise realizations, and plotted in Fig.~\ref{fig:estimator}(d). The bias follows the $\sim\sigma_W^{2}$ dotted trend line, and at excessive values the measured return is dominated by noise (see Appendix~\ref{app:score}). The noise amplitude used in the optimizations corresponds to $\Delta J\approx0.01$, which is small in comparison with the measured return $J\approx0.65$.

\section{Discussion}\label{sec:discussion}
In contrast to prior \textit{in situ} adjoint methods, training via PPGT does not require separate adjoint experiments per iteration. Instead, many forward experiments are run subject to intrinsic or instrumented diffusion. The gradient of every parameter is then read off from these forward trajectories alone, requiring neither adjoint excitation nor prepared states. The substrate need only run, fluctuate, and be \textit{judiciously} observed over the support of the score, according to Eq.~\eqref{eq:PPGT_gradient}.

Beyond the standard \textit{in situ} adjoint calibrations, e.g., calibration of $\partial_\theta\mu$~\cite{GWLK2025}, PPGT requires calibrating the summary statistics $Q$, which characterize the covariance of the noise processes, not their individual realizations. The gradient estimate is then resolved from averages over the measured fields according to Eqs.~\eqref{eq:adjoint_estimator} and \eqref{eq:PPGT_gradient}. Experimental feasibility of the approach, which relies on many trajectories per iteration, depends on both a measurement rate sufficient to record the time-domain field response and an adequate repetition rate of trajectories. The trajectory burden can be mitigated by using a baseline or, since the estimator variance scales as $\sigma_W^{-2}$ (Fig.~\ref{fig:estimator}(c)), instrumenting noise to the level the substrate and return bias tolerate (Fig.~\ref{fig:estimator}(d)). In particular, RF and microwave implementations (${\sim}100\,\mathrm{kHz}$--$10\,\mathrm{GHz}$ range) offer arguably the best timescale tradeoff: infrared or optical settings (${\sim}100\,\mathrm{MHz}$--$100\,\mathrm{THz}$) strain time-domain measurement rates, while mechanical or acoustic devices (${\sim}1\,\mathrm{Hz}$--$10\,\mathrm{kHz}$) may be limited by repetition rate.

\section{Conclusion}\label{sec:conclusion}
Noise inherent to physical processes, intrinsic or instrumented, is the resource that renders the gradient resolvable through measurement in PPGT. As the physical analog of PGT, it is primed to inherit the established concepts of reinforcement learning, such as partial observability~\cite{kaelbling1998planning}, continual learning~\cite{wang2024comprehensivesurveycontinuallearning}, and off-policy reuse with trust-region methods (e.g., TRPO, PPO)~\cite{lin1992self,schulman2017proximalpolicyoptimizationalgorithms,schulman2015trust}. Since PPGT is formulated at the generator level, it is applicable to general Markov processes (e.g., jump-diffusions), with only the score function changing form. Even non-Markovian extensions, such as colored noise, memory kernels, or spectrally filtered rewards, might be incorporated through state augmentation, provided white-noise diffusion still spans each parameter perturbation of the augmented dynamics.

On the practical side, the many trajectories required by PPGT \textit{in situ} can potentially be amortized for continuing, rather than episodic, tasks~\cite{Baxter_2001} (e.g., steady-state operation under continuous drive) by partitioning a single time-series measurement into multiple weakly correlated sub-trajectories. A proof-of-principle experimental demonstration in RF circuitry~\cite{GWLK2025,guillamon2026insituadjointwavecontrol,kwon2026insitutimedomainphysicaladjoint} is currently underway. Eventually, this \textit{in situ} design approach could be implemented via parallel device arrays, thereby naturally accounting for mesoscopic fluctuations and fabrication errors among ensemble members. We view PPGT as an early step toward physical, rather than digital, intelligence~\cite{hauser2012role,chen2026physical,wang2026embodying}, in which body and brain merge into a shared substrate---opening a path toward adaptive, self-optimizing, and ultimately, agentic metamaterials that autonomously execute multistep tasks~\cite{keshvari2026adaptivesensingnonadaptiveinformation}.

\begin{acknowledgments}
We thank Lautaro Moragues for discussions that motivated the treatment of degenerate diffusion. W.T. and Z.L. are supported by the U.S. Army Research Office (award numbers W911NF2410390 and W911NF2510113).
\end{acknowledgments}

\section*{Data availability}
The production scripts, figure drivers, and data files are openly available in Zenodo at \href{https://doi.org/10.5281/zenodo.23226207}{10.5281/zenodo.23226207}.

\section*{AI disclosure}
Anthropic's Claude was used under the authors' direction for literature and formulation review, GPU migration of the authors' CPU codebase, implementation of figure drivers, and manuscript revision. The authors verified all outputs and take full responsibility for this work.

\section*{Author contributions}
W.T. developed and implemented the theoretical framework and wrote the manuscript. Z.L. proposed the connection to the policy gradient theorem and guided the work.

\appendix

\section{Neglected Gradient Terms}\label{app:neglected}
The objective function in expectation value form can equivalently be expressed, $J(\theta)=\langle g_T,p(\,\cdot\,,T)\rangle+\int_{t_0}^Tdt\langle g(\,\cdot\,,t),p(\,\cdot\,,t)\rangle$, with the inner-product notation, $\langle u,v\rangle\equiv\int_\mathcal{X} dx\,u(x)\,v(x)$. The full adjoint gradient reads
\begin{align}\label{eq:raw_gradient}
\nabla_\theta J&=\left\langle\partial_\theta g_T\left(\,\cdot\,\right),p\left(\,\cdot\,,T\right)\right\rangle+\left\langle \lambda\left(\,\cdot\,,t_0\right),s_0\left(\,\cdot\,\right)\right\rangle\\
&+\int_{t_0}^Tdt\,
\left\langle\partial_\theta g\left(\,\cdot\,,t\right)+\left(\partial_\theta\mathcal{L}_t\lambda\right)\left(\,\cdot\,,t\right),p\left(\,\cdot\,,t\right)\right\rangle.\notag
\end{align}
The initial sensitivity, $s_0(x)\equiv\partial_\theta p_0(x;\theta)$, and the reward functions' parameter dependence usually vanish or can be obtained in a closed form; both are neglected in Eq.~\eqref{eq:reduced_gradient}.

When non-trivial, the initial sensitivity term in Eq.~\eqref{eq:raw_gradient}, $\langle \lambda(\,\cdot\,,t_0),s_0(\,\cdot\,)\rangle$, must also be amenable to sampling. The simplest scenario is when the initial distribution is known analytically, $p_0(x;\theta)=p(x,t_0;\theta)$. In this case, the log-trick obtains $\int dx'\,\lambda(x',t_0)\,\partial_\theta p_0=\int dx'\, p_0\,\lambda(x',t_0)\,\partial_\theta\ln p_0
=\mathbb{E}[\hat\lambda_0\,\partial_\theta\ln p_0]$, where we have replaced the initial adjoint with the sample-based return-to-go of Eq.~\eqref{eq:adjoint_estimator}. Including its associated baseline $\bar\lambda_s$ (see Appendix~\ref{app:baseline}), 
\begin{equation}\label{eq:baselined_caseA_IC_corr}
 \left\langle \lambda\left(\,\cdot\,,t_0\right),s_0\left(\,\cdot\,\right)\right\rangle=\mathbb{E}\left[\left(\hat\lambda_0-\bar\lambda_s\right)\partial_\theta\ln p_0\right].
\end{equation}
The initial sensitivity estimator can also be formulated using the reparameterization trick~\cite{mohamed2020monte, caflisch2024adjoint,SM}.

\section{Baseline Variance Reduction}\label{app:baseline}
A cheap and effective variance reduction method is to introduce a baseline, $\bar \lambda$, in sampled expressions involving a score. Assuming the baseline is independent from the score, $\mathbb{E}[(\hat\lambda-\bar \lambda)\,U_j]=\mathbb{E}[\hat\lambda \,U_j]-\bar \lambda\,\mathbb{E}[U_j]$, where $U_j$ denotes the score-function sampled along trajectory $j$. Since the expected value of a score is zero, the baseline does not bias the estimate. The baselined variance is
$\varsigma^2\equiv\mathbb{E}[(\hat\lambda-\bar \lambda)^2\,U_j^2]-\mathbb{E}[(\hat\lambda-\bar\lambda)\,U_j]^2=\varsigma_0^2+\bar\lambda^2\,\mathbb{E}[U_j^2]-2\bar\lambda\,\mathbb{E}[\hat\lambda\,U_j^2 ]$, where $\varsigma_0^2$ is the baseline-free variance. Differentiating with respect to the baseline, $\nabla_{\bar\lambda}\varsigma^2 = 2\bar\lambda\,\mathbb{E}[U_j^2]-2\,\mathbb{E}[\hat\lambda\,U_j^2]$, the variance minimizer is $\bar\lambda=\,\mathbb{E}[\hat\lambda\,U_j^2]/\mathbb{E}[U_j^2]$. To avoid correlation bias, the baseline per trajectory must be independent from the corresponding sampled score-function, which is achieved using leave-one-out expectations:
$\bar\lambda^{(j)}=\sum_{i\ne j}\hat\lambda^{(i)}\,U_i^2/\sum_{i\ne j}\,U_i^2$,
where $\hat\lambda^{(j)}$ indicates the return-to-go sampled along trajectory $j$, while trajectory-dependent $U_j$ and $\bar\lambda\rightarrow\bar\lambda^{(j)}$ are decorrelated by excluding the corresponding sample from the baseline estimate. When $\mathbb{E}[U_j^2]$ diverges, as can happen when nondegeneracy is supplied by multiplicative noise alone~\cite{SM}, convergence under adaptive step sizes is governed instead by a finite fractional moment of the gradient noise~\cite{zhang2020adaptive}, though the baseline above remains a practical default. Leave-one-out trajectory-index notation is suppressed elsewhere.

For further variance reduction, we condition on the state by partitioning the samples into $K=20$ quantile bins of the total field intensity on sites interacting with the corresponding parameter ($\lvert\psi_n\rvert^2$ for $\omega_n$ and $\chi_n$, $\lvert\psi_n\rvert^2+\lvert\psi_m\rvert^2$ for $\kappa_{nm}$), evaluated at the preceding step so that bin membership is independent of the noise increment. Leave-one-out expectations are applied within each bin to yield state-dependent, per-step baselines, which do not bias the estimate under the conditioning.

\section{Complete Stochastic-Adjoint Gradient Estimator}\label{app:complete}
Restoring the neglected gradient terms of Eq.~\eqref{eq:raw_gradient} to Eq.~\eqref{eq:PPGT_gradient} and assuming a closed form of the initial sensitivity as in Eq.~\eqref{eq:baselined_caseA_IC_corr}, the complete stochastic-adjoint gradient estimator is
\begin{align}\label{eq:PPGT_estimator}
&\mathcal{G}\big(\{X_{t_n}\}_{n=0}^{\mathcal{T}}\big)
=\left(\hat\lambda_0-\bar\lambda_s\right)\partial_\theta\ln p_0(X_{t_0})+\partial_\theta g_T(X_{t_\mathcal{T}})+\notag\\&\,\quad\Delta t\sum_{n=0}^{\mathcal{T}-1}\partial_\theta g(X_{t_n},t_n)
+\sum_{n=0}^{\mathcal{T}-1}\Delta\lambda_{n}U(X_{t_n},X_{t_{n+1}}),
\end{align}
where $\Delta\lambda_{n}\equiv\hat\lambda_{n+1}-\bar\lambda_{n}$. The first term is the initial sensitivity and the $\bar\lambda$ are baselines for variance reduction.

\section{Score with Diffusion Terms}\label{app:score}
When $\mu$ and $Q$ are both $\theta$-dependent, the per-step score of an It\^o process following Euler--Maruyama updates (discretized form of Eq.~\eqref{eq:main_ito}; $dt\rightarrow\Delta t$) takes the full form,
\begin{align}\label{eq:parameter_score}
U(x, x';\theta)=&-\frac{1}{2}\mathrm{Tr}\left[Q^{-1}\partial_\theta Q\right]\\+r\left(x,x'\right)^\top&Q^{-1}\left(\frac{1}{2}\partial_\theta\mu(x,t)+\frac{1}{4\Delta t}\partial_\theta Q\,Q^{-1}r\left(x,x'\right)\right),\notag
\end{align}
where $r(x,x')\equiv x'-x-\mu(x,t)\Delta t$. Evaluation of Eq.~\eqref{eq:parameter_score} requires $Q$ to be full rank, or at least $\partial_\theta\mu\in\mathrm{range}(Q)$ and $\mathrm{range}(\partial_\theta\sigma)\subseteq\mathrm{range}(Q)$. Here, we considered uniform radiative loss, enforcing full-rank $Q$. When diffusion is sparse, nondegeneracy can be restored \textit{in situ} by instrumenting noise along unnoised directions (e.g., via multiplicative noise, where additive noise injection is inconvenient), or by averaging the diffusion over a finite sampling interval, across which fast, conservative exchange redistributes fluctuation power among the directions it couples, which happens isotropically over both quadratures of a single resonator~\cite{SM}.

Furthermore, training parameters directly associated with noise channels, e.g., $\gamma_e$, can risk wireheading the reward. For example, growing $\gamma_e$ amplifies noise injected at the output port, which the chosen form of the return cannot distinguish from transmitted power from the input port, driving it above unity; see Fig.~\ref{fig:estimator}(d). Separately, the estimator variance for volatility-type parameters diverges as the time step is refined, a well-known limitation of likelihood-ratio methods with standard remedies~\cite{broadie1996estimating,glasserman2004monte}.

\section{Temporal Coupled Mode Theory}\label{app:tcmt}
The TCMT equations of motion are written in the form $i\ket*{\dot\psi}=H_\mathrm{eff}\ket{\psi}+iD(\ket{s^{+}}+\sigma^{(e)}\ket*{\dot{\eta}^{(e)}})+iD^{(r)}\sigma^{(r)}\ket*{\dot{\eta}^{(r)}}$ and $\ket{s^-}=-\ket{s^+}+D^\top\ket{\psi}$, where $\ket{\psi}\in\mathbb{C}^N$ and $\ket{s^{\pm}}\in\mathbb{C}^M$ are vectors of complex amplitudes that represent the $N$ resonators of the network and $M$ scattering channels, $D\in\mathbb{C}^{N\times M}$ ($D^{(r)}\in\mathbb{C}^{N\times N^{(r)}}$) is the coupling matrix that connects the accessible scattering channels ($N^{(r)}=N$ unobservable reservoirs) with the resonator network. For simplicity, the coupling strengths with each of the external channels, $\gamma_{e_m}$, are taken to be non-dispersive, here assumed to be symmetric and normalized $\gamma_e=\gamma_{e_1}=\gamma_{e_2}=1$, and encoded by the coupling matrix elements $\bra{n_m}D\ket{e_m}=\sqrt{2\gamma_{e_m}}$. Losses to radiative channels are described by the decay rates $\gamma_n$, here assumed to be uniform $\gamma_n=\gamma=0.02$, and are encoded by $\bra{n}D^{(r)}\ket{r_n}=\sqrt{2\gamma_n}$. The notation $\ket{n_m}$ indicates that lead $m$ is connected to resonator $n=n_m$, and $\ket{e_m}$ ($\ket{r_n}$) a basis element of the external (radiative) channels.

The dynamics are described by the effective Hamiltonian
\begin{align}\label{eq:Effective_Hamiltonian}
H_{\mathrm{eff}}(\theta,t)=&\sum_n \left[\left(\omega_n(\theta)+\chi_n(\theta)\vert\psi_n\vert^2\right)\ketbra{n}{n}\right]\\
+\sum_{m\ne n}&\left(\kappa_{nm}(\theta
)+\delta\kappa_{nm}(t)\right)\ketbra{n}{m}-i\Gamma\notag,
\end{align}
where $\ket{n}$ denotes the local orthonormal basis element associated with the $n^\mathrm{th}$ resonator mode with amplitude $\psi_n\equiv\braket{n}{\psi}$; $\Gamma\equiv(DD^\top+D^{(r)}D^{(r)\top})/2$ is the total decay-rate matrix due to external coupling and radiative loss channels. Open channels host additive noise through $\ket{\eta^{(e)}}\in\mathbb{C}^{M}$ and $\ket{\eta^{(r)}}\in\mathbb{C}^{N^{(r)}}$, whose quadratures are standard Wiener processes with their increments sampled from a zero-mean, unit-variance normal distribution, while $\sigma^{(e)}$ and $\sigma^{(r)}$ respectively encode the additive noise amplitudes associated with external and radiative channels, which are identically set to $\sigma_W=0.5$. The $N$ complex-variable TCMT equations can be transformed to a $2N$ real-variable representation equivalent to Eq.~\eqref{eq:main_ito}~\cite{SM}, where $H_\mathrm{eff}$, $\ket{\psi}$, $D$, and $\ket{s^+}$ enter through the drift, while $D$ and $D^{(r)}$ enter through the noise.

Time-dependent perturbations applied to the internal couplings were drawn independently per bond $(n,m)$ and trajectory $j$: $\delta\kappa_{nm}^{(j)}(t)=\sigma_\kappa\sum_{k=1}^{N_\mathrm{dyn}}\tilde{a}_{k,j}\sin(k\Omega\,t+\phi_{k,j})$, where $\tilde{a}_{k,j}=a_{k,j}/\sqrt{\sum_{k'} a_{k',j}^2}$ with $a_{k,j}\sim\mathcal{N}(0,1)$ so that $\sigma_\kappa$ alone sets the scale, $\phi_{k,j}\sim\mathcal{U}(0,2\pi)$, $N_\mathrm{dyn}=10$, and $\Omega=\Omega_\mathrm{max}/N_\mathrm{dyn}$ with $\Omega_\mathrm{max}=1$ (in units of the coupling scale). In contrast to Wiener processes whose white noise is resampled each step, this source of variance is correlated in time. The Fourier coefficients give a deterministic function of $t$ that enters through the drift rather than the diffusion.

The partial derivatives required to evaluate the score are readily obtained from Eq.~\eqref{eq:Effective_Hamiltonian}: $\partial_{\omega_n}H_\mathrm{eff}=\ketbra{n}{n}$, $\partial_{{\kappa_{nm}}}H_\mathrm{eff}=\ketbra{n}{m}+\ketbra{m}{n}$, and $\partial_{\chi_n} H_\mathrm{eff}=|\psi_n|^2\ketbra{n}{n}$. Inverse solves can be accelerated using the conjugate-gradient method~\cite{CGM,SM}, by computing the matrix-vector products $Qv$ in a closed form, though here $Q$ is diagonal and the solve reduces to a componentwise reciprocal. Other parameter types are also treated in~\cite{SM}.

\section{Optimization Details}\label{app:opt}
Every lattice was initialized with identical resonator frequencies $\omega_n=10$, uniform coupling $\kappa_{nm}=1$, and $\chi_n\sim\mathcal{N}(0,\sigma_\chi^2)$ with $\sigma_\chi=10^{-4}$. Euler--Maruyama updates were integrated with time step $\Delta t=10^{-3}$. Optimizations used a learning rate $\alpha=0.05$, normalized per component by the root mean square $h_l$ of the gradient estimate over the $\mathcal{E}$ trajectories of iteration $l$, denoted $\langle\cdot\rangle_{\mathcal{E}}$. To mitigate batch-to-batch fluctuations, $h_l$ was clamped within a factor $\rho=10$ of an exponential moving average $\bar h_l$ (decay $\beta=0.9$), $\theta_{l+1}=\Pi_\Theta[\theta_l+\alpha\,\langle\mathcal{G}(\theta_l)\rangle_{\mathcal{E}}/\tilde h_l]$ (all componentwise operations), where $\tilde h_l=\mathrm{clamp}(h_l,\bar h_l\rho^{-1},\bar h_l\rho)$, $\bar h_{l}=\beta \bar h_{l-1}+(1-\beta)h_l$ ($\bar h_1=h_1$), and $h_l=\sqrt{\langle\mathcal{G}(\theta_l)^2\rangle_{\mathcal{E}}}+10^{-8}$. Here, $\Pi_\Theta$ indicates projection to the constrained training space $\theta\in\Theta$.

The trajectory budget per iteration, $\mathcal{E}$, was increased in increments of $1{,}000$ up to $20{,}000$ and then in increments of $5{,}000$ up to a ceiling of $30{,}000$, advancing whenever the objective was detected to have stopped improving according to a trend-detection test~\cite{SM}. In practice, the trend-detection parameters were set too sensitively, which led to spurious over-escalation toward the budget ceiling: at the set learning rate, progress within the trend-detection window can be slow and stochastic enough that the trend test cannot reliably detect it.

Alternatively to the corridor membership, the relevance of each parameter in the trained configuration is indicated in Fig.~\ref{fig:convergence}(b--d), where curve opacity logarithmically encodes, within each family, the time-integrated field exposure at zero Wiener noise, $\int dt\,\langle f(X_t)\rangle_\mathcal{E}$, with $f(X_t)=\lvert\psi_n\rvert^2$ for $\omega_n$, $f(X_t)=\lvert\psi_n\psi_m\rvert$ for $\kappa_{nm}$, and $f(X_t)=\lvert\psi_n\rvert^4$ for $\chi_n$, where $\langle \cdot\rangle_\mathcal{E}$ denotes a sample average over $\mathcal{E}$ trajectories and $\psi_n$ is the complex field amplitude on site $n$.

\bibliography{bibtex}

\end{document}


\title{Supplemental Material:\\
Physical policy gradient theorem for \textit{in situ} stochastic-adjoint training}

\author{William Tuxbury}
\email{wtuxbury@vt.edu}
\affiliation{Bradley Department of Electrical and Computer Engineering,
Virginia Tech, Blacksburg, Virginia 24061, USA}

\author{Zin Lin}
\affiliation{Bradley Department of Electrical and Computer Engineering,
Virginia Tech, Blacksburg, Virginia 24061, USA}

\maketitle

The purpose of this supplementary material (SM) is to provide theoretical background and details for the derivation of the physical policy gradient theorem (PPGT). Additionally, this SM includes a stochastic temporal coupled mode theory (TCMT) formulation that incorporates both additive and multiplicative noise channels, with analytic partial derivatives and matrix-vector products required for the training parameters in the main text and others. The last section discusses further optimization details pertaining to the trajectory budget and baseline conditioning.

\section{Theoretical Framework}
\subsection{Stochastic Differential Equations}
Stochastic dynamics driven by white noise are commonly modeled as It\^o processes, where the state increment is composed of a deterministic drift and a random noise term that produces diffusive evolution, \begin{equation}\label{eq:ItonD}
dX_t=\mu(X_t,t)\,dt+\sigma(X_t,t)\,dW_t,
\end{equation}
where $X_t\in\mathcal{X}\subseteq\mathbb{R}^\mathcal{M}$ is the $\mathcal{M}$-dimensional state and $W_t\in\mathbb{R}^\mathcal{W}$ is a $\mathcal{W}$-dimensional vector of standard Wiener processes. Meanwhile, $\mu(x,t)\in\mathbb{R}^\mathcal{M}$ is the drift vector, $\sigma(x,t)$ is an $\mathcal{M}\times \mathcal{W}$ matrix that maps the Wiener processes to $\mathcal{X}$, and $Q(x,t)=\frac{1}{2}\sigma(x,t)\sigma(x,t)^\top$ is the $\mathcal{M}\times\mathcal{M}$ diffusion tensor. Dependency of $\mu(x,t)$ and $\sigma(x,t)$ on the parameter vector $\theta\in\mathbb{R}^\mathcal{P}$ is suppressed for brevity.

Stochastic differential equations (SDEs) can be numerically integrated using the Euler--Maruyama method. Given the drift and diffusion of the It\^o process, Eq.~\eqref{eq:ItonD}, together with the initial condition $X_{t_0}$ and time-step $\Delta t$, the stochastic process can be evolved following the recursion~\cite{kloeden1992numerical},
\begin{equation}\label{eq:EM_update}
X_{t_{n+1}}\approx \bar x(X_{t_n},t_n)+\sigma(X_{t_n},t_n)\,\Delta W_{t_n},\qquad\text{where}\quad \bar x(x,t)\equiv x+\mu(x,t)\Delta t.
\end{equation}
Here, $t_n\equiv t_0+n\Delta t$ for $n=0,\dots,\mathcal{T}$ with $t_\mathcal{T}=T$, and the recursion runs over $n=0,\dots,\mathcal{T}-1$. Each component of the Wiener increments, $\Delta W_{t_n}$, is drawn from a normal distribution with zero mean and variance equal to the time step, $\mathcal{N}(0,\Delta t)$. The approximation is exact in the limit $\Delta t\rightarrow0$.

\subsubsection{Fokker--Planck Correspondence}

The components of $W_t$ are independent Wiener processes with quadratic variation $dt$, giving the multiplication rules,
\begin{equation}\label{eq:ito_rules}
dt^2=0,\qquad dW_t^i\,dt=0,\qquad dW_t^i\,dW_t^j=\delta_{ij}\,dt.
\end{equation}
Throughout, superscripts denote the index of the vector component of random variables ($X_t$ and $W_t$) and otherwise subscripts denote the indices of vector and matrix components. Therefore, from Eq.~\eqref{eq:ItonD}, to first order in $dt$,
\begin{equation}\label{eq:Qvar}
dX_t^idX_t^j=\sum_{l=1}^\mathcal{W}\sum_{k=1}^\mathcal{W}\sigma_{il}\sigma_{jk}dW_t^ldW_t^k=\sum_{l=1}^\mathcal{W}\sigma_{il}\sigma_{jl}dt=2Q_{ij}\,dt.
\end{equation}
In the multi-variate Taylor expansion of $\varphi\in C_c^2(\mathbb{R}^\mathcal{M})$,
\begin{equation}
d\varphi(X_t)=\sum_{i=1}^\mathcal{M} \partial_i\varphi(X_t)dX_t^i+\frac{1}{2}\sum_{i=1}^\mathcal{M}\sum_{j=1}^\mathcal{M}\partial_{ij}\varphi(X_t)dX_t^idX_t^j,
\end{equation}
where $\partial_i\equiv\frac{\partial}{\partial x_i}$ and $\partial_{ij}\equiv\frac{\partial^2}{\partial x_i\partial x_j}$.
Using Eqs.~\eqref{eq:ItonD} and \eqref{eq:Qvar}, retaining terms first order in $dt$, we obtain It\^o's lemma~\cite{ito1951stochastic},
\begin{align}
d\varphi(X_t)&=\sum_{i=1}^\mathcal{M} \partial_i\varphi(X_t)\,\mu_i\!\left(X_t,t\right)\,dt+\sum_{i=1}^\mathcal{M}\sum_{j=1}^\mathcal{W} \partial_i\varphi(X_t)\,\sigma_{ij}\!\left(X_t,t\right)dW_t^j+\sum_{i=1}^\mathcal{M}\sum_{j=1}^\mathcal{M}\partial_{ij}\varphi(X_t)Q_{ij}\!\left(X_t,t\right)dt\notag\\
&=\left[\mu(X_t,t)\cdot\nabla_x \varphi(X_t)+\mathrm{Tr}\left(Q(X_t,t)\nabla_x^2\varphi(X_t)\right)\right]dt+\left[\sigma^\top\!(X_t,t)\nabla_x \varphi(X_t)\right]\cdot dW_t,\label{eq:dg_w_ito}
\end{align}
where $\nabla_x^2$ denotes the Hessian such that $[\nabla_x^2\varphi]_{ij}\equiv\partial_{ij}\varphi$, and we have used the fact that $Q$ is symmetric by its construction. The above result can be used to obtain Dynkin's formula~\cite{dynkin2012theory},
\begin{align}
\frac{d}{dt}\mathbb{E}\left[\varphi(X_t)\right]&=\mathbb{E}\left[(\mathcal{L}_t\varphi)(X_t)\right]=\int_\mathcal{X}dx\,(\mathcal{L}_t\varphi)(x)p(x,t)=\langle(\mathcal{L}_t\varphi)(\,\cdot\,),p(\,\cdot\,,t)\rangle,\label{eq:Dynkin}\\
\text{where}\quad(\mathcal{L}_t\varphi)(x)&\equiv\mu(x,t)\cdot\nabla_x \varphi(x)+\mathrm{Tr}\left[Q(x,t)\nabla_x^2\varphi(x)\right].\notag
\end{align}
Here, $\mathcal{L}_t$ is the backward generator, $p(x,t)$ is the probability density, and $\langle u,v\rangle\equiv\int_\mathcal{X} dx\,u(x)\,v(x)$ indicates the inner product, with $(\,\cdot\,)$ the argument position of the integration variable. The expectation value of the last term in Eq.~\eqref{eq:dg_w_ito} vanishes by the martingale property and we assume standard regularity of $\varphi$ to commute expectation value with the derivative. Additionally, for any operator $\mathcal{A}$, its adjoint $\mathcal{A}^\dagger$ is defined such that $\langle u,\mathcal{A}v\rangle=\langle \mathcal{A}^\dagger u,v\rangle$
for any $u,v:\mathcal{X}\rightarrow\mathbb{R}$ where the inner product exists, which gives the forward generator $\mathcal{L}_t^\dagger$ as the adjoint of the backward generator,
\begin{equation}\label{eq:adjoint_generator_def}
\langle(\mathcal{L}_t\varphi)(\,\cdot\,),p(\,\cdot\,,t)\rangle=\langle\varphi(\,\cdot\,),(\mathcal{L}_t^\dagger p)(\,\cdot\,,t)\rangle,\qquad\forall\varphi\in C_c^2(\mathcal{X}).
\end{equation}

Since $\varphi$ is arbitrary, with this definition and Eq.~\eqref{eq:Dynkin} we immediately obtain the Fokker--Planck equation (FPE)~\cite{risken1989fokker},
\begin{equation}\label{eq:FPE_nD}
\partial_t p(x,t)=(\mathcal{L}_t^\dagger p)(x,t),\qquad p(x,t_0)=p_0(x),
\end{equation}
where $p_0(x)$ is the initial distribution and $\partial_t\equiv\frac{\partial}{\partial t}$. Integrating by parts in Eq.~\eqref{eq:adjoint_generator_def} to identify the form of the adjoint generator,
\begin{equation}
\left\langle(\mathcal{L}_t\varphi)(\,\cdot\,),p(\,\cdot\,,t)\right\rangle=\sum_{i=1}^\mathcal{M}\left\langle\partial_i\varphi(\,\cdot\,),\mu_i(\,\cdot\,,t)p(\,\cdot\,,t)\right\rangle+\sum_{i=1}^\mathcal{M}\sum_{j=1}^\mathcal{M}\left\langle \partial_{ij}\varphi(\,\cdot\,),Q_{ij}(\,\cdot\,,t)p(\,\cdot\,,t)\right\rangle.
\end{equation}
For the first terms,
\begin{equation}
\left\langle\partial_i\varphi(\,\cdot\,),\mu_i(\,\cdot\,,t)p(\,\cdot\,,t)\right\rangle=0-\left\langle\varphi(\,\cdot\,),\partial_i\!\left(\mu_i(\,\cdot\,,t)p(\,\cdot\,,t)\right)\right\rangle
\end{equation}
and for the second terms,
\begin{equation}
\left\langle\partial_{ij}\varphi(\,\cdot\,),Q_{ij}(\,\cdot\,,t)p(\,\cdot\,,t)\right\rangle=0-\left\langle\partial_{j}\varphi(\,\cdot\,),\partial_i\!\left(Q_{ij}(\,\cdot\,,t)p(\,\cdot\,,t)\right)\right\rangle=0+\left\langle\varphi(\,\cdot\,),\partial_{ij}\!\left(Q_{ij}(\,\cdot\,,t)p(\,\cdot\,,t)\right)\right\rangle.
\end{equation}
Compared with Eq.~\eqref{eq:adjoint_generator_def}, we have
\begin{equation}
(\mathcal{L}_t^\dagger p)(x,t)=-\nabla_x\cdot\left(\mu(x,t)p(x,t)\right)+\nabla_x\cdot\nabla_x\cdot\left(Q(x,t)\,p(x,t)\right).
\end{equation}
Then, subject to the initial condition $p(x,t_0)=p_0(x)$, Eq.~\eqref{eq:FPE_nD} is called the Kolmogorov forward equation. Using the product rule $\nabla_x\cdot(Q p)= Q\nabla_x p\,+p\nabla_x \cdot Q$ the FPE can be written in continuity form:
\begin{equation}\label{eq:FPE_flux_form}
\partial_tp(x,t)+\nabla_x\cdot\mathcal{J}(x,t)=0,\qquad p(x,t_0)=p_0(x),
\end{equation}
where $\mathcal{J}(x,t)\equiv\left[\mu(x,t)-\nabla_x\cdot Q(x,t)\right]p(x,t)-Q(x,t)\nabla_x p(x,t)$ is the probability flux and $[\nabla_x\cdot Q]_i=\sum_{j}\partial_jQ_{ij}$ is unambiguous because $Q$ is symmetric by construction.

\subsubsection{Noise-Induced Drift}
For multiplicative noise, $\sigma$ depends explicitly on the state, which can vary over the integration step. For extrinsic noise with finite correlation time, $\sigma$ must be evaluated along the step rather than at its start, which contributes a physical correction to the drift. The Stratonovich formulation~\cite{gardiner2009stochastic,wongzakai1965relation},
\begin{equation}
dX_t=\mu^\circ(X_t,t)\,dt+\sigma(X_t,t)\circ dW_t,
\end{equation}
where $\mu^\circ$ is the drift underlying the physical model, accounts for this by weighting the Wiener increments by $\sigma$ evaluated at the midpoint state, $X_t+\tfrac{1}{2}dX_t$, indicated by $\circ$. Under the Stratonovich convention, It\^o's lemma simplifies,
\begin{equation}
d\varphi(X_t)=\nabla_x\varphi(X_t)\circ dX_t,
\end{equation}
with $\circ$ contracting componentwise, which is the ordinary chain rule. The drift correction to the It\^o process follows from expanding the Stratonovich noise increment about the beginning of the step, retaining terms to $\mathcal{O}(dt)$:
\begin{align}
\sigma(X_t,t)\circ dW_t&\equiv\sigma\!\left(X_t+\tfrac{1}{2}dX_t,t\right)d W_t\approx\sum_{j=1}^\mathcal{W}\left[\sigma_{:j}\left(X_t,t\right)d W_t^j+\tfrac{1}{2}\nabla_x\sigma_{:j}\vert_{x=X_t}d X_t\,dW_t^j\right]\\
&=\sum_{j=1}^\mathcal{W}\left[\sigma_{:j}\left(X_t,t\right)dW_t^j+\tfrac{1}{2}\nabla_x\sigma_{:j}\vert_{x=X_t}\left(\mu(X_t,t)\,dt\,dW_t^j+\sum_{w=1}^\mathcal{W}\sigma_{:w}(X_t,t)\,dW_t^w\,dW_t^j\right)\right],\notag
\end{align}
where $\sigma_{:w}$ indicates the $w^\mathrm{th}$ column of $\sigma$, $\nabla_x\sigma_{:w}$ its Jacobian with the derivative index appended, $[\nabla_x\sigma_{:w}]_{i\ell}=\partial_\ell\sigma_{iw}$, and in the last equality we used Eq.~\eqref{eq:ItonD}. From Eq.~\eqref{eq:ito_rules}, the drift term vanishes and the $w$-sum collapses, giving the corrected It\^o drift,
\begin{equation}\label{eq:corrected_drift}
\mu(X_t,t)=\mu^\circ(X_t,t)+\tfrac{1}{2}\sum_{w=1}^\mathcal{W}\nabla_x\sigma_{:w}\vert_{x=X_t} \sigma_{:w}(X_t,t).
\end{equation}
For additive noise, $\nabla_x\sigma=0$, there is no drift correction and both formulations coincide, as in the main text, where only additive noise is retained.

\subsection{Physical Policy Gradient Theorem Formulation}

Define an objective function as an expectation value involving both terminal $g_T(X_T)$ and continuing $g(X_t,t)$ rewards,
\begin{equation}\label{eq:return}
J\!\left(\theta\right)=\mathbb{E}\left[g_T\left(X_T\right)+\int_{t_0}^Tdt\,g\left(X_t,t\right)\right]=\left\langle g_T,p(\,\cdot\,,T)\right\rangle+\int_{t_0}^Tdt\left\langle g,p\right\rangle,
\end{equation}
where in the second equality the expectation values are expressed as integrals over the distribution that solves the corresponding FPE, Eq.~\eqref{eq:FPE_nD}. The goal is to efficiently compute the gradient $\nabla_\theta J$ by sampling the original SDE.
\subsubsection{Adjoint Gradient}
Define the sensitivity $s(x,t)\equiv\partial_\theta p(x,t)$, where $\partial_\theta\equiv\frac{\partial}{\partial \theta}$ acts componentwise for each of the $\mathcal{P}$ parameters, and differentiate the FPE to obtain the sensitivity partial differential equation (PDE),
\begin{equation}
\partial_t s(x,t)-(\mathcal{L}_t^\dagger s)(x,t)-(\partial_\theta\mathcal{L}_t^\dagger p)(x,t)=0,\qquad s(x,t_0)=s_0(x),
\end{equation}
where $s_0(x)$ is the initial sensitivity. Multiply by a function $\lambda(x,t)$ and integrate-by-parts over time and space,
\begin{align}
0&=\int_{t_0}^Tdt\,\left\langle\lambda,\partial_t s\right\rangle-\int_{t_0}^Tdt\,\left\langle\lambda,\mathcal{L}_t^\dagger s\right\rangle-\int_{t_0}^Tdt\,\left\langle\lambda,\partial_\theta\mathcal{L}_t^\dagger p\right\rangle\\&
=\left\langle\lambda(\,\cdot\,,T),s(\,\cdot\,,T)\right\rangle-\left\langle\lambda(\,\cdot\,,t_0),s_0(\,\cdot\,)\right\rangle-\int_{t_0}^Tdt\,\left\langle\partial_t\lambda,s\right\rangle-\int_{t_0}^Tdt\,\left\langle\mathcal{L}_t\lambda,s\right\rangle-\int_{t_0}^Tdt\,\left\langle\partial_\theta\mathcal{L}_t\lambda, p\right\rangle\notag,
\end{align}
where arguments inside the temporal integrands are suppressed. Therefore,
\begin{equation}\label{eq:Adjoint_Identity}
\left\langle\lambda(\,\cdot\,,T),s(\,\cdot\,,T)\right\rangle-\int_{t_0}^Tdt\,\left\langle\partial_t\lambda,s\right\rangle-\int_{t_0}^Tdt\,\left\langle\mathcal{L}_t\lambda,s\right\rangle=\left\langle\lambda(\,\cdot\,,t_0),s_0(\,\cdot\,)\right\rangle+\int_{t_0}^Tdt\,\left\langle\partial_\theta\mathcal{L}_t\lambda, p\right\rangle.
\end{equation}

Now, with the necessary prerequisites established, we are ready to differentiate the objective function,
\begin{equation}
\nabla_\theta J=\left\langle\partial_\theta g_T(\,\cdot\,),p(\,\cdot\,,T)\right\rangle+\int_{t_0}^Tdt\,\left\langle\partial_\theta g,p\right\rangle+\left\langle g_T(\,\cdot\,),s(\,\cdot\,,T)\right\rangle+\int_{t_0}^Tdt\,\left\langle g,s\right\rangle.
\end{equation}
At this point, we can choose the $\lambda$ to satisfy the following adjoint PDE and terminal condition~\cite{li2004adjoint}, also called the Kolmogorov backward equation,
\begin{equation}\label{eq:Adjoint_PDE}
\partial_t \lambda(x,t)+\left(\mathcal{L}_t\lambda\right)(x,t)+g(x,t)=0,\qquad\lambda\left(x,T\right)=g_T(x),
\end{equation}
where $g_T(x)$ is the terminal adjoint condition. Then, from Eq.~\eqref{eq:Adjoint_Identity}, we have
\begin{equation}
\left\langle g_T(\,\cdot\,),s(\,\cdot\,,T)\right\rangle+\int_{t_0}^Tdt\,\left\langle g,s\right\rangle=\left\langle \lambda(\,\cdot\,,t_0),s_0(\,\cdot\,)\right\rangle+\int_{t_0}^Tdt\,
\left\langle\partial_\theta\mathcal{L}_t\lambda,p\right\rangle,
\end{equation}
which we can use to eliminate intermediate sensitivities from the last two terms of the gradient expression. Thus,
\begin{equation}\label{eq:supp_raw_gradient}
\nabla_\theta J=\left\langle\partial_\theta g_T(\,\cdot\,),p(\,\cdot\,,T)\right\rangle+\left\langle \lambda(\,\cdot\,,t_0),s_0(\,\cdot\,)\right\rangle+\int_{t_0}^Tdt\,
\left\langle\partial_\theta g(\,\cdot\,,t)+(\partial_\theta\mathcal{L}_t\lambda)(\,\cdot\,,t),p(\,\cdot\,,t)\right\rangle,
\end{equation}
for which we note the middle term vanishes when the initial conditions of the original SDE are $\theta$-independent. In this case, it can be written as the expectation value,
\begin{equation}\label{eq:gradient_expectation}
\nabla_\theta J=\mathbb{E}\left[\partial_\theta g_T\left(X_T\right)+\int_{t_0}^T dt\,\partial_\theta g\left(X_t,t\right)+\left(\partial_\theta\mathcal{L}_t\lambda\right)\left(X_t,t\right)\right],
\end{equation}
which is the full form of Eq.~\eqref{M-eq:reduced_gradient} with arguments restored, appearing as its density form in Eq.~\eqref{M-eq:raw_gradient} of Appendix~\ref{M-app:neglected}.

\subsubsection{Feynman--Kac Formula}

The solution to the adjoint PDE Eq.~\eqref{eq:Adjoint_PDE} is amenable to sampling via the original SDE by the Feynman--Kac formula~\cite{shreve1991brownian}. Applying It\^o's lemma, Eq.~\eqref{eq:dg_w_ito}, to $\lambda(X_t,t)$,
\begin{align}\label{eq:Ito_Adjoint}
d\lambda(X_t,t)&=\left[\partial_t\lambda(X_t,t)+\left(\mathcal{L}_t\lambda\right)\left(X_t,t\right)\right]dt+\sigma^\top\!\left(X_t,t\right)\nabla_x\lambda\left(X_t,t\right)\cdot dW_t\notag\\
&=-g\left(X_t,t\right)\,dt+\sigma^\top\!\left(X_t,t\right)\nabla_x\lambda\left(X_t,t\right)\cdot dW_t,
\end{align}
where we used the form of the adjoint PDE, Eq.~\eqref{eq:Adjoint_PDE} (assuming the solution exists). Then, define the process
\begin{equation}
Y_u=\lambda\!\left(X_u,u\right)+\int_t^ud\tau\,g(X_\tau,\tau).
\end{equation}
Using the terminal condition, $\lambda(x,T)=g_T(x)$, the process at the boundary times $u=t$ and $u=T$ is $Y_t=\lambda(X_t,t)$ and $Y_T=g_T(X_T)+\int_t^Td\tau\,g(X_\tau,\tau)$, respectively. Differentiating, we have
\begin{equation}
dY_u=d\lambda(X_u,u)+g(X_u,u)du=\sigma^\top\!(X_u,u)\nabla_x\lambda(X_u,u)\cdot dW_u,
\end{equation}
where in the second equality we have inserted Eq.~\eqref{eq:Ito_Adjoint}. Hence, $Y_u$ is a martingale, $Y_t=\mathbb{E}\left[Y_T\big| X_t=x\right]$. Equivalently,
\begin{equation}\label{eq:FK}
\lambda(x,t)=\mathbb{E}\left[g_T\left(X_T\right)+\int_t^Td\tau\,g(X_\tau,\tau)\bigg\vert X_t=x\right],
\end{equation}
where we have used the condition at boundary time $u=t$. Therefore, $\lambda(x,t)$ can be sampled from the same It\^o process as that under stochastic optimization. This result is a simplified form of the well-known Feynman--Kac formula. Equations~\eqref{eq:return} and \eqref{eq:FK} give the standard identity $J=\left\langle p_0(\,\cdot\,),\lambda(\,\cdot\,,t_0)\right\rangle$, which reduces to $J=\lambda(x_0,t_0)$ for a deterministic initial condition. However, straightforward as it appears, evaluating it exhaustively is infeasible. It requires simulating an ensemble of $\mathcal{E}$ trajectories launched from every possible state and time. With $\mathcal{T}$ the number of discrete time steps and $x\in\mathbb{R}^\mathcal{M}$ discretized into $\mathcal{D}$ pixels per coordinate, the calculation scales exponentially with the state dimension, $\mathcal{O}\left(\mathcal{E}\mathcal{T}\mathcal{D}^\mathcal{M}\right)$. The only escape is to accept sparse sampling. Yet the gradient estimator involves state-space derivatives of the adjoint field through the differential operator $\mathcal{L}_t$, which pointwise estimates do not supply without interpolation from dense sampling. This requirement is circumvented by applying the log-trick to Dynkin's formula, Eq.~\eqref{eq:Dynkin}, giving Eq.~\eqref{M-eq:Markov_derivative}.

\subsubsection{Closed-Form Score Function}
Since the Wiener increment, $\Delta W_{t_n}$, in Eq.~\eqref{eq:EM_update} is Gaussian with zero mean and variance $\Delta t$, the transition kernel is also Gaussian. Therefore, it is determined by its mean $\mathbb{E}[X_{t_{n+1}}\vert X_{t_n}=x]=\bar x(x,t_n)$ and covariance $\Sigma(x,t_n)\equiv\mathbb{E}[X_{t_{n+1}}X_{t_{n+1}}^\top\vert X_{t_n}=x]-\bar x(x,t_n)\bar x(x,t_n)^\top=\sigma(x,t_n)\sigma(x,t_n)^\top\Delta t$, assumed positive definite. When it is not, the kernel is supported on the affine subspace $\bar x(x,t_n)+\mathrm{range}(\sigma)$, and the score exists only for parameters that leave that support invariant. This condition is considered in depth below. One way to derive the transition kernel is by transforming coordinates to the covariance eigenbasis. To this end, introduce the decomposition and basis transformation
\begin{equation}
\Sigma=\mathcal{U}\Lambda \mathcal{U}^\top,\qquad z=\mathcal{U}^\top\!\left[x'-\bar x(x,t)\right]\in\mathbb{R}^\mathcal{M},
\end{equation}
where $\mathcal{U}\mathcal{U}^\top=\mathbb{I}_\mathcal{M}$, $\Lambda=\mathrm{diag}(\varsigma_1,\varsigma_2,\dots,\varsigma_\mathcal{M})$, and note that $\det(\Sigma)=\det(\Lambda)=\prod_k\varsigma_k$. In this uncorrelated basis,
\begin{align}
p_\theta^{\Delta t}\left(z\vert x\right)&=\prod_{k=1}^{\mathcal{M}}\frac{1}{\sqrt{2\pi\varsigma_k}}\exp\left(-\frac{z_k^2}{2\varsigma_k}\right)=\frac{1}{\sqrt{(2\pi)^\mathcal{M}\det(\Sigma)}}\exp\left(-\frac{z^\top\Lambda^{-1}z}{2}\right).
\end{align}
Since $\mathcal{U}^\top$ is orthogonal, the Jacobian determinant has unit magnitude. Hence, the transition kernel in the original basis reads
\begin{equation}
p_\theta^{\Delta t}(x'\vert x)=\frac{1}{\sqrt{(2\pi)^\mathcal{M}\det\left[\Sigma(x,t)\right]}}\exp\left(-\frac{\left[x'-\bar x(x,t)\right]^\top\Sigma^{-1}(x,t)\left[x'-\bar x(x,t)\right]}{2}\right).
\end{equation}
All that remains is to differentiate the log kernel,
\begin{align}
\ln\left[p_\theta^{\Delta t}(x'\vert x)\right]=-&\frac{\mathcal{M}}{2}\ln\left(2\pi\right)-\frac{1}{2}\ln\left[\det\left(\Sigma\right)\right]-\frac{1}{2}\left(x'-\bar x\right)^\top\Sigma^{-1}\left(x'-\bar x\right),
\end{align}
where the arguments of $\Sigma=\Sigma(x,t)$ and $\bar x=\bar x(x,t)$ are suppressed. Hence, we have the score
\begin{equation}\label{eq:first_score}
\partial_\theta\ln\left[p_\theta^{\Delta t}(x'\vert x)\right]=-\frac{1}{2}\mathrm{Tr}\left[\Sigma^{-1}\partial_\theta\Sigma\right]+\frac{1}{2}\partial_\theta \bar x^\top\Sigma^{-1}r+\frac{1}{2}r^\top\Sigma^{-1}\partial_\theta\Sigma\,\Sigma^{-1}r+\frac{1}{2}r^\top\Sigma^{-1}\partial_\theta\bar x,
\end{equation}
where $r\equiv x'-\bar x(x,t)$ is the residual from the deterministic update. Since $\Sigma$ is symmetric and the terms are scalar, $\partial_\theta\bar x^\top\Sigma^{-1}r=(\partial_\theta\bar x^\top\Sigma ^{-1}\,r)^\top=r^\top\Sigma ^{-1}\partial_\theta\bar x$. So, we have
\begin{align}
\partial_\theta\ln\left[p_\theta^{\Delta t}(x'\vert x)\right]&=-\frac{1}{2}\mathrm{Tr}\left[\Sigma^{-1}\partial_\theta\Sigma\right]+r^\top\Sigma^{-1}\partial_\theta\bar x+\frac{1}{2}r^\top\Sigma^{-1}\partial_\theta\Sigma\,\Sigma^{-1}r.
\end{align}
In terms of the drift $\mu(x,t)$ and diffusion $Q(x,t)\equiv\frac{1}{2}\sigma(x,t)\sigma(x,t)^\top$, for which $\Sigma(x,t)=2\Delta tQ(x,t)$, we obtain Eq.~\eqref{M-eq:parameter_score}.

In a later section we analyze the initial sensitivity term in Eqs.~\eqref{M-eq:raw_gradient} and \eqref{eq:supp_raw_gradient}, where it will be useful to have the score with respect to the initial state. From the Markov property, the path likelihood is
\begin{equation}
p_\theta^{\Delta t}\left(X_{t_\mathcal{T}},\dots,X_{t_1}\vert X_{t_0}\right)=p_\theta^{\Delta t}\left(X_{t_1}\vert X_{t_0}\right)\prod_{n=1}^{\mathcal{T}-1}p_\theta^{\Delta t}\left(X_{t_{n+1}}\vert X_{t_n}\right).
\end{equation}
Only the first-step likelihood depends on the initial state, therefore similar to Eq.~\eqref{M-eq:parameter_score}, we have
\begin{align}\label{eq:IC_score}
\nabla_{x}&\ln p_\theta^{\Delta t}\left(X_{t_\mathcal{T}},\dots,X_{t_1}\vert x\right)\big\vert_{x=X_{t_0}}=\nabla_{x}\ln p_\theta^{\Delta t}\left(X_{t_1}\vert x\right)\big\vert_{x=X_{t_0}}=-\frac{1}{2}\mathrm{Tr}\left[Q^{-1}\nabla_x Q\big\vert_{x=X_{t_0}}\right]\\
&+\left(X_{t_1}-X_{t_0}-\mu\left(X_{t_0},t_0\right)\Delta t\right)^\top Q^{-1}\left(\frac{\mathbb{I}_\mathcal{M}}{2\Delta t}+\frac{1}{2}\nabla_x \mu\big\vert_{x=X_{t_0}}+\frac{1}{4\Delta t}\nabla_x Q\big\vert_{x=X_{t_0}}Q^{-1}\left(X_{t_1}-X_{t_0}-\mu\left(X_{t_0},t_0\right)\Delta t\right)\right),\notag
\end{align}
where $\mathbb{I}_\mathcal{M}$ is the $\mathcal{M}\times\mathcal{M}$ identity matrix, arising from the Jacobian $\nabla_x\bar x=\mathbb{I}_\mathcal{M}+\Delta t\nabla_x\mu$.

In the purely deterministic case, $\sigma=0$, trajectories no longer explore their neighborhood, and score-based Monte Carlo estimation breaks down because the score of $p_\theta^{\Delta t}(x'\vert x)=\delta\left(x'-\bar{x}(x,t)\right)$ is undefined. The gradient itself remains well-defined, since the generator in Eq.~\eqref{eq:Dynkin} gives $\partial_\theta\mathcal{L}_t\lambda\rightarrow\partial_\theta\mu\cdot\nabla_x\lambda$, recovering the pathwise adjoint gradient along the noiseless trajectory~\cite{TDadjoint}. Sparse, but nonzero, diffusion is treated in the following sections.

\subsubsection{Inverse-Free Score Evaluation}

The closed-form expressions for the score function involve the diffusion tensor inverse, $Q^{-1}$, which does not exist when $\mathrm{rank}(\sigma)<\mathcal{M}$. However, $Q^{-1}v$ can be understood as the solution $u$ of $Qu=v$ rather than as an explicit inverse, which for singular $Q$ and $v\in\mathrm{range}(Q)$ is the unique Moore--Penrose pseudo-inverse solution. Even when $Q^{-1}$ exists, it remains advisable to avoid materializing either $Q$ or its inverse. Instead, the trace term can be refactored,
\begin{equation}\label{eq:trace_term}
\mathrm{Tr}\left[Q^{-1}\partial_\theta Q\right]=\mathrm{Tr}\left[\sigma^\top Q^{-1}\partial_\theta \sigma\right]=\sum_{w=1}^{\mathcal{W}}\sigma_{:w}^\top Q^{-1}\partial_\theta\sigma_{:w}.
\end{equation}
The first equality follows from the product rule, giving $Q^{-1}\partial_\theta Q=\tfrac{1}{2}Q^{-1}[\left(\partial_\theta \sigma\right)\sigma^\top+ \sigma\left(\partial_\theta\sigma\right)^\top]$. The $1/2$ factor cancels because both terms have the same trace, $\mathrm{Tr}[\sigma^\top Q^{-1}\partial_\theta\sigma]$, by cyclic trace permutation together with $Q$ being real symmetric and the invariance of the trace under transpose. Evaluating the score therefore requires solving $Qu=r$ and $Qu=\partial_\theta\sigma_{:w}$, which can be obtained by the conjugate gradient method~\cite{CGM} provided the right-hand side lies in $\mathrm{range}(Q)=\mathrm{range}(\sigma)$. For the residual, this holds by construction, since from Eq.~\eqref{eq:EM_update} $r\equiv X_{t_{n+1}}-\bar x(X_{t_n},t_n)=\sigma\,\Delta W_{t_n}\in\mathrm{range}(\sigma)$. Meanwhile, it holds for the columns of the noise partial when $\partial_\theta\sigma_{:w}\in\mathrm{range}(\sigma)$, which is the case away from channel openings or closures (e.g., vanishing coupling). With a zero initial guess, the conjugate gradient iterates stay in $\mathrm{range}(Q)$ up to roundoff. 

However, since $Q^{-1}r\in\mathrm{range}(Q)$, only the component of $\partial_\theta\mu$ in that subspace contributes to the contraction, so a parameter acting outside $\mathrm{range}(Q)$ is invisible to the score. When the noise is multiplicative, $Q$ may approach degeneracy near particular states. Exactly degenerate states have zero measure along stochastic trajectories, so the projection they induce contributes negligible bias. However, trajectories in their proximity can inflate the estimator variance, potentially to the point of heavy tails. Nondegenerate diffusion must reach every direction in which a trained parameter perturbs the dynamics over a kernel transition. Physically, fluctuation is the resource the score measures against, without which there is no stochastic variation along the perturbation for the estimator to register. Recovering nondegeneracy under sparse diffusion is considered below. Alternate \textit{in situ} approaches that exploit deterministic structure in the degenerate subspace may also be possible. Explicit matrix-vector products for the diffusion and its partial derivatives, along with those of the drift, are derived in the Relevant Quantities for the PPGT section. In the main text the noise is additive and uncorrelated, so $Q$ is diagonal and each solve reduces to a componentwise reciprocal.

\subsubsection{Coarse Graining Sparse Noise}

Consider the It\^o process given by
\begin{equation}\label{eq:sparse_noise_model}
du=f(u,v,t)\,dt+\sigma\,dW_t,\qquad
dv=h(u,v,t;\theta)\,dt,
\end{equation}
with the full state represented $x\doteq\begin{pmatrix}u&v\end{pmatrix}^\top$. Such structure arises when noise channels couple to some physical variables but not others, as for noise driving velocity but not position~\cite{kolmogorov1934zufallige}, or in a circuit implementation where Wiener processes drive the voltages but not the currents. The stepwise residual only populates entries corresponding to $u$, $r\doteq\begin{pmatrix}r_u&0\end{pmatrix}^\top$. Meanwhile, the $u$ block of the diffusion tensor is $Q_u=\sigma\sigma^\top/2$, but the $v$ block is zero, so that while $y=\begin{pmatrix}Q_u^{-1}r_u&0\end{pmatrix}^\top$ solves $Qy=r$, the evaluated one-step score is identically zero $\partial_\theta \ln p_\theta^{\Delta t}\propto y^\top\partial_\theta\mu=0$ and the parameter sensitivity is blind. This is because PPGT resolves the sensitivities through their competitions with fluctuations, which locally vanish when conditioned over a single step. One scheme to recover blind sensitivities is to marginalize the kernel over $k$ time steps, allowing the fluctuations to propagate to $v$ provided $\partial_uh\ne0$. Observing the process every $k^\mathrm{th}$ step leaves a Markov chain, so Eqs.~\eqref{M-eq:reduced_gradient}--\eqref{M-eq:PPGT_gradient} apply unchanged with $\Delta t\rightarrow k\Delta t$; only the kernel is replaced, $p_\theta^{\Delta t}(x'\vert x)\rightarrow p_\theta^{k\Delta t}(x''\vert x)$. The marginalized kernel is approximated as Gaussian by linearizing the Euler--Maruyama update about a reference trajectory $\check{x}_{n+i}$ that depends only on the kernel arguments $x=X_{t_n}$, $x''=X_{t_{n+k}}$, and the sub-step index $0\leq i\leq k$,
\begin{equation}\label{eq:delta_recursion}
\delta_{n+i+1}=\delta_{n+i}+\left[\mu(x_{n+i},t_{n+i})-\mu(\check x_{n+i},t_{n+i})\right]\Delta t+\sigma(x_{n+i},t_{n+i})\Delta W_{t_{n+i}}\approx\Phi_{n+i}\delta_{n+i}+\check \sigma_{i+1}\Delta W_{t_{n+i}},
\end{equation}
with $\check \sigma_{i+1}\equiv\sigma(\check x_{n+i},t_{n+i})$ and $\Phi_{n+i}\equiv\mathbb{I}_\mathcal{M}+\nabla_x\mu(x,t_{n+i})\vert_{x=\check x_{n+i}}\Delta t$ evaluated along the reference path for $0\leq i\leq k-1$. The drift state gradient, $\nabla_x\mu(x,t)$, is given in Relevant Quantities for the PPGT. Equation~\eqref{eq:delta_recursion} unrolls to $\delta_{n+2}=\Phi_{n+1}\check\sigma_1\Delta W_{t_n}+\check\sigma_2 \Delta W_{t_{n+1}}$, $\delta_{n+3}=\Phi_{n+2}\Phi_{n+1}\check\sigma_1\Delta W_{t_n}+\Phi_{n+2}\check\sigma_2 \Delta W_{t_{n+1}}+\check\sigma_3 \Delta W_{t_{n+2}}$, and in general,
\begin{equation}\label{eq:coarse_delta}
\delta_{n+k}=\sum_{i=1}^{k}\Psi_i\,\check\sigma_i\,\Delta W_{t_{n+i-1}},
\end{equation}
with the ordered product
$\Psi_i\equiv\Phi_{n+k-1}\Phi_{n+k-2}\cdots\Phi_{n+i}$, where $0\leq i\leq k$ indexes substeps within the window beginning at $n$, with $\Psi_k=\mathbb{I}_\mathcal{M}$. Equation~\eqref{eq:coarse_delta} is a linear functional of Wiener increments and therefore Gaussian. Since the Wiener increments are i.i.d., $\mathbb{E}[\Delta W_{t_{i}}\Delta W_{t_{j}}^\top]=\mathbb{I}_\mathcal{W}\Delta t$ for $i=j$ and zero otherwise, the linearized $k$-step diffusion tensor is
\begin{equation}\label{eq:coarse_Q}
Q_k\equiv\frac{1}{2k\Delta t}\mathbb{E}[\delta_{n+k}\delta_{n+k}^\top]=\frac{1}{2k}\sum_{i=1}^{k}\Psi_i\,\check\sigma_i\check\sigma_i^\top\,\Psi_i^\top.
\end{equation}
Each term in the sum carries the rank of $\check\sigma_i\check\sigma_i^\top$, but their ranges are rotated to $\Psi_i\,\mathrm{range}(\check\sigma_i)$, so the sum spans their union, $\mathrm{span}\{\Psi_1\check\sigma_1,\Psi_2\check\sigma_2,\dots,\Psi_{k}\check\sigma_k\}$, which can extend the support to $v$ when $\partial_u h\neq0$. Fluctuations reach $v$ only by accumulating through $h$, so they arrive integrated rather than injected. Comparable variance requires a full coupling time, $k\Delta t\sim 1/\|\partial_u h\|$, where $\|\partial_u h\|\equiv\min_i\inf_{\eta\in\zeta_i}\left\vert\partial_u h\vert_{\check{x}_{n+i}}\eta\right\vert$ bounds the propagation rate of fluctuations. The infimum is taken over $\zeta_i=\{\eta\,:\,\lvert\eta\rvert=1,\;\eta\perp\ker(\partial_u h\vert_{\check{x}_{n+i}})\}$, unit perturbations in $u$ that couple into $v$ at substep $i$.

If the reference trajectory $\check{x}_{n+i}$ is taken as the noiseless trajectory from $x$, $\check x_{n+i+1}=\check x_{n+i}+\mu(\check x_{n+i},t_{n+i})\Delta t$, then Eq.~\eqref{eq:delta_recursion} is a small-noise expansion~\cite{gardiner2009stochastic}. So, under the linearization $\check x_{n+k}$ is the mean of the marginalized kernel. Hence, the linearized parameter sensitivity obeys
\begin{equation}\label{eq:drift_recursion}
\partial_\theta\check x_{n+i+1}=
\partial_\theta\check x_{n+i}+
\left[\nabla_x\mu\vert_{\check x_{n+i}}\partial_\theta\check x_{n+i}+
\partial_\theta\mu(\check x_{n+i},t_{n+i})\right]\Delta t=
\Phi_{n+i}\partial_\theta\check x_{n+i}+
\partial_\theta\mu(\check x_{n+i},t_{n+i})\Delta t.
\end{equation}
Unrolling the Eq.~\eqref{eq:drift_recursion} recursion similar to Eq.~\eqref{eq:coarse_delta}, $\partial_\theta\check x_{n+k}=k\Delta t\,\mu_\theta ^{(k)}$, with
\begin{equation}\label{eq:coarse_drift_partial}
 \mu_\theta ^{(k)}\equiv\frac{1}{k}\sum_{i=1}^{k}\Psi_i\,\partial_\theta\mu(\check x_{n+i-1},t_{n+i-1}).
\end{equation}
The approximation is exact for linear drift with state-independent noise. Otherwise it degrades as the window lengthens, imposing an upper bound on $k\Delta t$ that cannot be relieved by shortening the window, since that sacrifices the very dispersal coarse-graining was introduced to achieve. As established above, the covariance contributed at each substep is $\check\sigma_i\check\sigma_i^\top\Delta t$. Assuming the propagation neither appreciably amplifies nor attenuates the fluctuations over the window, $\Psi_i^\top\Psi_i\approx\mathbb{I}_\mathcal{M}$, and since contributions from the independent increments of the $k$ substeps add in quadrature, Eq.~\eqref{eq:coarse_delta} bounds the mean-square deviation accumulated by step $k$, $\mathbb{E}[\delta_{n+k}^\top\delta_{n+k}]\lesssim\|\sigma\|^2\mathcal{M}k\Delta t$, with $\|\sigma\|\equiv\max_i\sup_{\vert\eta\vert=1}\vert\check\sigma_i^\top\,\eta\vert$ the maximal noise amplitude over substeps and directions. The step size enters only through the window $k\Delta t$, so refining $\Delta t$ at fixed window does not improve the expansion. The upper bound follows from the neglected second-order term, $\tfrac12\big\vert\delta_{n+i}^\top\nabla_x^2\mu|_{\check x_{n+i}}\delta_{n+i}\big\vert$, against the retained first-order term in Eq.~\eqref{eq:delta_recursion},
\begin{equation}\label{eq:coarse_grained_bound_1}
\frac{\|\nabla_x^2\mu\|}{\|\nabla_x\mu\|}\|\sigma\|\sqrt{k\Delta t}\ll1,
\end{equation}
where $\|\nabla_x^2\mu\|\equiv\max_i\sup_{\vert\eta\vert=1}\left\vert\eta^\top\nabla_x^2\mu\vert_{\check{x}_{n+i}}\,\eta\right\vert$ is the maximal curvature with respect to state and $\|\nabla_x\mu\|^2\equiv\max_i\frac{1}{\mathcal{M}}\mathrm{Tr}[\nabla_x\mu^\top\vert_{\check{x}_{n+i}}\nabla_x\mu\vert_{\check{x}_{n+i}}]$ is the normalized Frobenius norm of the drift state gradient as an isotropic scaling estimate, whose $\mathcal{M}^{-1/2}$ cancels the dimensional factor in the accumulated root-mean-square deviation. The lower bound of the approximation is determined by the required steps for fluctuations to propagate to $v$. Over $k$ substeps, $\mathrm{Var}(u)\sim \|\sigma\|^2\,k\Delta t$, due to the perturbation $\sim\|\sigma\|\sqrt{\Delta t}$ at step $n$, which couples to $v$ at the rate $\|\partial_uh\|\|\sigma\|\sqrt{\Delta t}$ for the remaining $k-i$ substeps, each displacing $\|\partial_uh\|\|\sigma\|\sqrt{\Delta t}(k-i)\Delta t$. Therefore, the inherited variance is
\begin{equation}
\mathrm{Var}(v)\sim\sum_{i=1}^k\left[\|\partial_uh\|\|\sigma\|\sqrt{\Delta t}(k-i)\Delta t\right]^2\sim \frac{1}{3}\left(\|\partial_uh\|k\Delta t\right)^2\|\sigma\|^2k\Delta t.
\end{equation}
Requiring comparable fluctuations in the unnoised subspace, $\mathrm{Var}(v)\sim\mathrm{Var}(u)$, gives $\|\partial_uh\|\,k\Delta t\sim1$. Together with Eq.~\eqref{eq:coarse_grained_bound_1}, the approximation must satisfy
\begin{equation}\label{eq:coarse_approx_bounds}
\frac{1}{\|\partial_uh\|}\sim
k\Delta t
\ll\left(\frac{\|\nabla_x\mu\|}{\|\sigma\|\|\nabla_x^2\mu\|}\right)^2.
\end{equation}
The definitions of normed quantities were set to conservatively estimate timescales bounding the window. Practically, the lower bound is set by the coupling time to disperse fluctuations, the upper by the time at which their accumulation invalidates the linearization, and the two are compatible when $\|\sigma\|^2\|\nabla_x^2\mu\|^2\ll\|\partial_uh\|\|\nabla_x\mu\|^2$. However, Eq.~\eqref{eq:coarse_approx_bounds} neglects dissipation, which damps fluctuations before they can propagate to $v$ over the window. The coupling must therefore also outrun the fastest decay rate over the window, $\gamma_\star\lesssim\|\partial_uh\|$, with $\gamma_\star\equiv \max_i\max\{-\Re[\mathrm{spec} (\nabla_x\mu\vert_{\check{x}_{n+i}})]\}$.

The coarse-grained $Q_k$ replaces $Q$, $\partial_\theta Q_k$ replaces $\partial_\theta Q$, the window-averaged sensitivity $\mu_\theta ^{(k)}$ replaces $\partial_\theta\mu$, and the residual, $r=x''-\check x_{n+k}$, is taken over the window in Eq.~\eqref{M-eq:parameter_score} with $\Delta t\rightarrow k\Delta t$. Evaluating the coarse-grained score requires $\Psi_i$ and $\partial_\theta\Psi_i$, since $\theta$ alters how fluctuations propagate through $\Phi_i$, even when $\partial_\theta\check\sigma_i=0$. Corresponding substitutions apply in Eq.~\eqref{eq:IC_score}, where the one-step Jacobian $\nabla_x\bar x=\Phi_0$ is replaced by the full window propagator $\Psi_0$ from $t_0$ to $t_{k}$. The return-to-go is likewise accumulated on the coarse grid, since finer rewards carry sensitivity the coarsened score does not resolve. However, this scheme covertly smuggles a model of the dynamics back into the estimator via the linearized propagators. In simulation, where that model is directly available, the scheme mainly serves as a model-based reference. In the following section, the same formulation will prove useful \textit{in situ}: taken to the continuous-time limit at finite sampling interval, it yields an effective diffusion with rank recovered under timescale separation.

\subsubsection{Effective Diffusion Under Timescale Separation}

In the continuous-time limit, $\Delta t\rightarrow0$, with the sampling interval $\tau\equiv k\Delta t$ held fixed, the summation in Eq.~\eqref{eq:coarse_Q} becomes an integral over $s\in[t_n,t_n+\tau]$, giving the window-averaged effective diffusion tensor (cf.\ the controllability Gramian~\cite{kalman1963controllability}),
\begin{equation}\label{eq:Q_eff_gen}
Q_k\rightarrow Q_\mathrm{eff}(x,t_n)\equiv\frac{1}{\tau}\int_{t_n}^{t_n+\tau}ds\,\Psi(t_n+\tau,s)\,Q\left(\check x(s),s\right)\,\Psi(t_n+\tau,s)^\top,
\end{equation}
with $\Psi_i\rightarrow\Psi(t_n+\tau,s)$, the continuum propagator. Analogously to Eq.~\eqref{eq:coarse_delta}, it obeys the recursion relation $\Psi(t+dt,s)=\Phi(t)\Psi(t,s)$, giving the evolution operator equation,
\begin{equation}\label{eq:operator_eq}
\partial_t\Psi(t,s)=A(t)\Psi(t,s),\qquad\Psi(s,s)=\mathbb{I}_\mathcal{M}
\end{equation}
for the infinitesimal propagator $\Phi(t)\equiv\mathbb{I}_\mathcal{M}+A(t)\,dt$, with generator $A(t)\equiv\nabla_x\mu(x,t)\vert_{x=\check x(t)}$ evaluated along the noiseless reference path. For a time-independent generator, the solution is $\Psi(t,s)=\exp[A\,(t-s)]$. Substituting $s\rightarrow t_n+\tau-s$ over the sampling interval,
\begin{equation}\label{eq:Q_eff_const_A}
Q_\mathrm{eff}(x,t_n)=\frac{1}{\tau}\int_0^\tau ds\,
\exp[A\,s]\,Q\big(\check x(t_n+\tau-s),\,t_n+\tau-s\big)\,\exp[A^\top\,s].
\end{equation}
When the dynamics and sampling interval admit the timescale separation~\cite{pavliotis2008multiscale}, $\tau_\omega\lesssim \tau\ll\tau_\gamma$, the generator can be decomposed into a constant, real-valued fast part $A_\omega$ and a slow part $A_\gamma(t)$ that carries all state and time dependence along the reference path, $A(t)=A_\omega+A_\gamma(t)$. In the rotating frame, the fast timescale evolution factors out exactly, $\Psi(t,s)=e^{A_\omega(t-s)}\Psi_\gamma(t,s)$, so from Eq.~\eqref{eq:operator_eq} the remaining propagator $\Psi_\gamma$ obeys $\partial_t\Psi_\gamma=e^{-A_\omega(t-s)}A_\gamma\,e^{A_\omega(t-s)}\Psi_\gamma$. If the fast exchange conserves energy, $A_\omega$ is diagonalizable with a purely imaginary spectrum. Therefore, $\exp[{A_\omega(t-s)}]$ is purely oscillatory and thus bounded. Hence, with $A_\gamma=\mathcal{O}(\tau_\gamma^{-1})$, we have $\Psi_\gamma=\mathbb{I}_\mathcal{M}+\mathcal{O}(\tau/\tau_\gamma)$~\cite{gronwall1919note} and $\Psi(t,s)=e^{A_\omega(t-s)}+\mathcal{O}(\tau/\tau_\gamma)$ over the sampling interval and from Eq.~\eqref{eq:Q_eff_const_A},
\begin{equation}\label{eq:Qeff_fastavg}
Q_\mathrm{eff}(x,t_n)=\frac{1}{\tau}\int_0^\tau ds\,Q_\omega(s)+\mathcal{O}\!\left(\frac{\tau}{\tau_\gamma}\right),
\end{equation}
where $Q_\omega(s)\equiv \exp[A_\omega s]\,Q\big(\check x(t_n+\tau-s),\,t_n+\tau-s\big)\,\exp[A_\omega^\top s]$. Since $A_\omega:T_x^*\mathcal{X}\times T_x\mathcal{X}\rightarrow\mathbb{R}$, where $T_x\mathcal{X}$ and $T_x^*\mathcal{X}$ denote the tangent and cotangent space at $x$, respectively, the generator can be restricted to each invariant block $p\in\{1,\dots,N_\omega\}$ as
\begin{equation}
\Pi_p\,A_\omega\,\Pi_p=\sum_{j\in\mathcal{P}_p}i\Omega_j\,P_j,\qquad\Pi_p\equiv\sum_{j\in\mathcal{P}_p}P_j,
\end{equation}
where $\sum_{p=1}^{N_\omega}\Pi_p=\mathbb{I}_\mathcal{M}$ partitions the space. The $P_j$ are rank-one, complex, and in general oblique eigenprojectors and the $\Omega_j$ are the corresponding eigenfrequencies, with $\mathcal{P}_p$ indexing block $p$. Assuming $A_\omega$ is a real-valued construction, its eigenfrequencies come in conjugate pairs, $\Omega_{\bar j}=-\Omega_j$, with $P_{\bar j}=\overline{P_j}$, or as self-paired zeros with real eigenprojectors. Likewise, since $Q:T_x^*\mathcal{X}\times T_x^*\mathcal{X}\rightarrow\mathbb{R}$, taking $Q$ to vary on the slow timescale along the reference path, as when it depends on the state only through its slowly varying envelope, from Eq.~\eqref{eq:Qeff_fastavg} the corresponding effective diffusion tensor block is
\begin{align}\label{eq:Qeff_block}
\Pi_p\,Q_\mathrm{eff}(x,t_n)\,\Pi_p^\top&=\frac{1}{\tau}\int_0^\tau ds\sum_{j,j'\in\mathcal{P}_p}e^{i(\Omega_j+\Omega_{j'})s}\,P_j\,Q\big(\check x(t_n+\tau-s),\,t_n+\tau-s\big)\,P_{j'}^\top\notag\\
&=\sum_{j\in\mathcal{P}_p}P_j\,Q(x,t_n)\,P_j^\dagger+\mathcal{O}\!\left(\frac{\tau}{\tau_\gamma}\right)+\mathcal{O}\!\left(\frac{1}{\lvert\Omega_j+\Omega_{j'}\rvert\,\tau}\right)_{\Omega_{j'}\neq-\Omega_j},
\end{align}
where the last term is the residual of the non-conjugate cross terms, which dephase over the sampling interval. Degenerate eigenvectors are chosen such that $P_jQP_{j'}^\dagger=0$ for $j\neq j'$ with $\Omega_j=\Omega_{j'}$. Assuming the noise is not correlated between blocks, $\Pi_p\,Q\,\Pi_{p'}^\top=0$ for $p\ne p'$, or they are non-resonant on the scale of the sampling interval, the full effective diffusion tensor can be approximately assembled,
\begin{equation}
Q_\mathrm{eff}\approx\sum_{p=1}^{N_\omega}\Pi_p\,Q_\mathrm{eff}\,\Pi_p^\top.
\end{equation}

A useful example is to consider one block with two degrees of freedom, only one of which is noised, as for a resonator with only one Wiener-driven quadrature~\cite{roberts1986stochastic}. This is Eq.~\eqref{eq:sparse_noise_model} with scalar $u$, $v$, and $Q_u$, so that, suppressing the block index, $\Pi\,Q\,\Pi^\top=\mathrm{diag}(Q_u,0)$. In energy-normalized coordinates, the restricted fast-exchange generator, with conjugate eigenfrequencies $\Omega_\pm=\pm\tau_\omega^{-1}\ne0$ and circular eigenvectors, is represented
\begin{equation}
\Pi\,A_\omega\,\Pi\doteq\tau_\omega^{-1}\begin{pmatrix}0&-1\\1&0\end{pmatrix},\qquad P_\pm\doteq\frac{1}{2}\begin{pmatrix}1&\pm i\\\mp i&1\end{pmatrix},
\end{equation}
with $P_\pm$ its eigenprojectors. Therefore, from Eq.~\eqref{eq:Qeff_block},
\begin{equation}\label{eq:res_Qeff}
\Pi\,Q_\mathrm{eff}\,\Pi^\top\approx P_+QP_+^\dagger+P_-QP_-^\dagger=\frac{Q_u}{2}\mathbb{I}_2,
\end{equation}
which leaves the block trace invariant. Physically, by Dynkin's formula, Eq.~\eqref{eq:Dynkin} applied to $\varphi=\lvert\delta\rvert^2$, the squared deviation from the noiseless reference path in energy-normalized coordinates, the fluctuation power injected by the noise, $2\,\mathrm{Tr}(Q)$, is conserved after the window average and only redistributed within the block. Importantly, Eqs.~\eqref{eq:res_Qeff} and, more generally, \eqref{eq:Qeff_block} depend only on the eigenstructure of the fast exchange generator in each block, enabling \textit{in situ} admissible recovery of the diffusion rank through the choice of sampling interval. Conversely, if the timescale separation is satisfied at the sampling interval, scoring with the unaveraged $Q=\mathrm{diag}(Q_u,0)$ biases the estimator. Moreover, each block can be measured directly as $\Pi_p\,Q_\mathrm{eff}\,\Pi_p^\top\approx\tfrac{1}{2\tau}\,\mathrm{Cov}(\Pi_p \tilde r)$. Here, the tilde denotes the residual taken in the rotating frame over the sampling interval $\tau$, in which the fast exchange is demodulated and the nearly constant remainder of the drift is evaluated at the measured state $X_{t_n}$.

\subsubsection{Reparameterization Trick in the Initial Conditions}

Alternatively to the analytic initial conditions considered in Appendix~\ref{M-app:neglected} of the main text, the initial distribution can be reparameterized such that initial state $X_{t_0}$ absorbs the parameter dependency of the distribution enabling its random variables to be drawn from a fixed distribution,
\begin{equation}
X_{t_0}=\mathcal{R}\!\left(\xi,\theta\right),\qquad\text{where}\quad\xi\sim P_\xi~\left(\theta\,\text{-independent}\right).
\end{equation}
Since $\lambda(x,t_0)$ is not an explicit function of $\theta$, by Leibniz rule:
\begin{align}\label{eq:repar_initsens}
\left\langle \lambda\left(\,\cdot\,,t_0\right),s_0\left(\,\cdot\,\right)\right\rangle&=\int dx' \lambda\!\left(x',t_0\right)\frac{\partial p(x',t_0)}{\partial\theta}
=\frac{\partial }{\partial\theta}\int dx' \lambda\!\left(x',t_0\right)p(x',t_0)\\
&=\frac{\partial}{\partial\theta}\int d\xi\,\lambda\!\left(\mathcal{R}\!\left(\xi,\theta\right),t_0\right)p_\xi=\frac{\partial}{\partial\theta}\mathbb{E}_\xi\left[\lambda\!\left(\mathcal{R}\!\left(\xi,\theta\right),t_0\right)\right]=\mathbb{E}_\xi\left[\Upsilon^\top\frac{\partial}{\partial\theta}\mathcal{R}\!\left(\xi,\theta\right)\right],\notag\\
\text{where}\quad
\Upsilon&\equiv\nabla_x\lambda(x,t_0)\vert_{x=X_{t_0}}.\notag
\end{align}
To evaluate this term, separate the initial reward and express the return-to-go over the pathwise marginal probabilities,
\begin{align}
\lambda\!\left(X_{t_0},t_0\right)=&\lim_{\Delta t\rightarrow0}\Delta t\,g\!\left(X_{t_0},t_0\right)+\mathbb{E}\left[\lambda\!\left(X_{t_1},t_1\right)\right]\\
=&\lim_{\Delta t\rightarrow0}\Delta t\,g\!\left(X_{t_0},t_0\right)+
\int dX_{t_\mathcal{T}}\dots dX_{t_2}\,dX_{t_1}\lambda\!\left(X_{t_1},t_1\right)p_\theta^{\Delta t}\!\left(X_{t_\mathcal{T}},\dots,X_{t_1}\vert X_{t_0}\right)\notag.
\end{align}
Computing the gradient using Eq.~\eqref{eq:IC_score} and applying the log-trick again,
\begin{align}
\nabla_x\lambda\left(x,t_0\right)\vert_{x=X_{t_0}}
&=\lim_{\Delta t\rightarrow0}\Delta t\,\nabla_{x}g\left(x,t_0\right)\vert_{x=X_{t_0}}\notag\\
&\quad+\int dX_{t_\mathcal{T}}\dots dX_{t_2}\,dX_{t_1}\,p_\theta^{\Delta t}\!\left(X_{t_\mathcal{T}},\dots,X_{t_1}\vert x=X_{t_0}\right)\lambda\!\left(X_{t_1},t_1\right)\nabla_{X_{t_0}}\ln p_\theta^{\Delta t}\!\left(X_{t_1}\vert x=X_{t_0}\right),
\end{align}
which is approximated at finite $\Delta t$ by the expectation
\begin{equation}\label{eq:sample_init_adjoint_state_grad}
\Upsilon\approx\Delta t\,\nabla_{x}g\!\left(x,t_0\right)\vert_{x=X_{t_0}}+\mathbb{E}\left[\lambda\!\left(X_{t_1},t_1\right)\nabla_{x}\ln p_\theta^{\Delta t}\!\left(X_{t_1}\vert x\right)\vert_{x=X_{t_0}}\right].
\end{equation}
The leading finite-$\Delta t$ correction, which vanishes in the strict continuum limit, is retained for completeness at finite time step. Since $\Upsilon$ is resolved from the first transition alone, Eq.~\eqref{eq:sample_init_adjoint_state_grad} requires $Q$ to be nondegenerate at $X_{t_0}$. When $Q$ degenerates only at particular states, as for multiplicative noise, this fails on a set of zero measure for $\xi$-dependent $\mathcal{R}$, whereas $\xi$-independent $\mathcal{R}$ must avoid those states exactly. Moreover, the estimate inherits the term $r^\top Q^{-1}/(2\Delta t)$ from Eq.~\eqref{eq:IC_score}, which is $\mathcal{O}(1/\sqrt{\Delta t})$ since $r\sim\sigma\sqrt{\Delta t}$ and $Q^{-1}\sim\sigma^{-2}$. Hence, $\mathrm{Var}(\Upsilon)\sim1/\Delta t$ shares the pathology of volatility-type parameters, except that it does not accumulate over the trajectory. However, since $\nabla_x\ln p_\theta^{\Delta t}$ is a score, the estimation of $\Upsilon$ admits baseline variance reduction (see Appendix~\ref{M-app:baseline} of the main text).

\newpage

\section{Stochastic Temporal Coupled Mode Theory}
\subsection{Equations of Motion}
The TCMT equations of motion can be written as~\cite{joannopoulos2008photonic, zhu2013temporal}
\begin{align}
i\frac{d}{dt}\ket{\psi}&=\left(H_\mathrm{eff}+H^{(s)}\right)\ket{\psi}+iD\left(\ket{s^{+}}+\Theta^{(e)}\ket*{\dot{\eta}^{(e)}}\right)+iD^{(r)}\Theta^{(r)}\ket*{\dot{\eta}^{(r)}},\notag\\
\ket{s^-}&=-\ket{s^+}+D^\top\ket{\psi},\label{eq:CMT}
\end{align}
where $\ket{\psi}\in\mathbb{C}^N$ and $\ket{s^{\pm}}\in\mathbb{C}^M$. The coupling matrix $D\in\mathbb{C}^{N\times M}$ connects the $M$ controllable scattering channels with the $N$-resonator network and is assumed non-dispersive (Markovian), while $D^{(r)}\in\mathbb{C}^{N\times N^{(r)}}$ connects the $N^{(r)}$ unobservable reservoirs, representing internal loss. Together, these open channels carry additive noise through $\ket{\eta^{(e)}}\in\mathbb{C}^{M}$ and $\ket{\eta^{(r)}}\in\mathbb{C}^{N^{(r)}}$, whose quadratures are standard Wiener processes with increments drawn from a zero-mean, unit-variance normal distribution. The amplitudes are encoded in $\Theta^{(e)}=\mathrm{diag}(\sigma^{(e)}_1,\dots,\sigma^{(e)}_{M})$ and $\Theta^{(r)}=\mathrm{diag}(\sigma^{(r)}_1,\dots,\sigma^{(r)}_{N^{(r)}})$. In the main text these are uniform within each family and equal across them, $\sigma_W\equiv\sigma^{(e)}_m=\sigma^{(r)}_n$ for all $m$ and $n$, so $\Theta^{(e)}=\sigma_W\mathbb{I}_M$ and $\Theta^{(r)}=\sigma_W\mathbb{I}_{N^{(r)}}$, with $N^{(r)}=N$. The wave operator is written as a sum of two terms. The first term is the deterministic part of the effective Hamiltonian,
\begin{equation}
H_\mathrm{eff}=H_0(t)+\sum_nf_n\left(\lvert\braket{n}{\psi}\rvert^2\right)\ketbra{n}{n}-\frac{i}{2}\left(DD^\top+D^{(r)}D^{(r)\top}\right),
\end{equation}
where $H_0(t)\in\mathbb{C}^{N\times N}$ is Hermitian at every instant, $f_n$ describe local intensity-dependent shifts to the resonator frequencies, and $\ket{n}$ is the local orthonormal basis associated with the $n^\mathrm{th}$ resonator mode. For modeling simplicity, open nonlinear channels are not considered, since they would induce nonlinear Langevin terms in Eq.~\eqref{eq:CMT} by the fluctuation--dissipation relation. The coupling to unobservable reservoirs is assumed constant and non-dispersive, so the dynamics remain Markovian. The second term is the stochastic part of the effective Hamiltonian and encodes multiplicative noise (e.g., parameter jitter),
\begin{equation}
H^{(s)}=\sum_{j=1}^{N^{(s)}} \sigma_j^{(s)}\xi^{(s)}_j(t)H_j,
\end{equation}
where $\xi^{(s)}_j(t)=\dot W^{(s)}_j$ is the (distributional) time derivative of standard Wiener processes with $\Delta W^{(s)}_j\sim\mathcal{N}(0,\Delta t)$. Noise enters through the Hermitian $H_j$, a shape perturbation whose scale is set by $\sigma_j^{(s)}\in\mathbb{R}$. The product $H^{(s)}\ket{\psi}$ in Eq.~\eqref{eq:CMT}, as well as any $H^{(s)}$ state dependence, is interpreted in the Stratonovich sense, following the Noise-Induced Drift section. Multiplicative noise is absent in the main text, $N^{(s)}=0$. Later we will also make use of another orthonormal basis $\ket*{j^{(s)}}$ to represent a vector of these Wiener processes as $\ket*{W^{(s)}}=\sum_{j=1}^{N^{(s)}}W^{(s)}_j\ket*{j^{(s)}}$, such that $W_j^{(s)}=\braket*{j^{(s)}}{W^{(s)}}$.
\subsection{Transform to Real-Variable Representation}
The variables and coefficients in Eq.~\eqref{eq:CMT} are generically complex. Therefore, before mapping them to an It\^o process, we will transform the equations of motion to be entirely real. To this end, we introduce the stacked vectors and operators,
\begin{align}
\ket*{\tilde{\psi}}\,&\dot\equiv\,
\begin{pmatrix}\ket*{\psi}\\ \ket*{\psi}^*\end{pmatrix},
&\qquad
\tilde{M}_\mathrm{eff}\,&\dot\equiv\,
\begin{pmatrix}M_\mathrm{eff}&0\\ 0&(M_\mathrm{eff})^*\end{pmatrix},
&\qquad
\tilde{D}\,&\dot\equiv\,
\begin{pmatrix}D&0\\ 0&D^*\end{pmatrix},\notag\\
\ket*{\tilde{s}^\pm}\,&\dot\equiv\,
\begin{pmatrix}\ket*{s^\pm}\\ \ket*{s^\pm}^*\end{pmatrix},
&\qquad
\tilde{M}^{(s)}\,&\dot\equiv\,
\begin{pmatrix}M^{(s)}&0\\ 0&(M^{(s)})^*\end{pmatrix},
&\qquad
\tilde{D}^{(r)}\,&\dot\equiv\,
\begin{pmatrix}D^{(r)}&0\\ 0&(D^{(r)})^*\end{pmatrix},\notag\\
\ket*{\tilde{\eta}^{(e)}}\,&\dot\equiv\,
\begin{pmatrix}\ket*{\eta^{(e)}}\\ \ket*{\eta^{(e)}}^*\end{pmatrix},
&\qquad
\tilde{\Theta}^{(e)}\,&\dot\equiv\,
\begin{pmatrix}\Theta^{(e)}&0\\ 0&\Theta^{(e)*}\end{pmatrix},
&\qquad
\tilde{\Theta}^{(r)}\,&\dot\equiv\,
\begin{pmatrix}\Theta^{(r)}&0\\ 0&\Theta^{(r)*}\end{pmatrix},\notag\\
\ket*{\tilde{\eta}^{(r)}}\,&\dot\equiv\,
\begin{pmatrix}\ket*{\eta^{(r)}}\\ \ket*{\eta^{(r)}}^*\end{pmatrix},
&\qquad
\text{where}\quad M_{\mathrm{eff}}\,&\equiv\,iH_{\mathrm{eff}},
&\qquad
\text{and}\quad M^{(s)}\,&\equiv\,iH^{(s)}.
\end{align}
In the stacked complex representation, $\ket{n}$ and $\ket{N+n}$ label the upper and lower blocks associated with resonator $n$. Equation~\eqref{eq:CMT}, together with its conjugate, can be written
\begin{equation}\label{eq:stacked_CMT}
\frac{d}{dt}\ket{\tilde{\psi}}=-\tilde{M}_{\mathrm{eff}}\ket{\tilde{\psi}}+\tilde{D}\ket{\tilde{s}^+}-\tilde{M}^{(s)}\ket{\tilde{\psi}}
+\tilde{D}\tilde{\Theta}^{(e)}\ket{\tilde{\eta}^{(e)}}+\tilde{D}^{(r)}\tilde{\Theta}^{(r)}\ket{\tilde{\eta}^{(r)}}.
\end{equation}
By introducing the following transformations,
\begin{equation}\label{eq:basis_transform}
V\equiv\begin{pmatrix}
\mathbb{I}_N&i\mathbb{I}_N\\
\mathbb{I}_N&-i\mathbb{I}_N
\end{pmatrix},\qquad
V^{(e)}\equiv\begin{pmatrix}
\mathbb{I}_M&i\mathbb{I}_M\\
\mathbb{I}_M&-i\mathbb{I}_M
\end{pmatrix},\qquad
V^{(r)}\equiv\begin{pmatrix}
\mathbb{I}_{N^{(r)}}&i\mathbb{I}_{N^{(r)}}\\
\mathbb{I}_{N^{(r)}}&-i\mathbb{I}_{N^{(r)}}
\end{pmatrix},
\end{equation}
we can transform the stacked complex-conjugate representation to a stacked real/imaginary representation,
\begin{align}
\ket{x}&\doteq
\begin{pmatrix}
\Re(\ket{\psi})\\
\Im(\ket{\psi})
\end{pmatrix},
&
\ket{b^\pm}&\doteq
\begin{pmatrix}
\Re(\ket{s^{\pm}})\\
\Im(\ket{s^{\pm}})
\end{pmatrix},
\\
\ket*{W^{(e)}}&\doteq
\begin{pmatrix}
\Re(\ket{\eta^{(e)}})\\
\Im(\ket{\eta^{(e)}})
\end{pmatrix},&\ket*{W^{(r)}}&\doteq
\begin{pmatrix}
\Re(\ket{\eta^{(r)}})\\
\Im(\ket{\eta^{(r)}})
\end{pmatrix}.\notag
\end{align}
An overbar on a ket indicates a standard basis vector of the real-variable representation, $\ket{\bar{k}}$ for $k=1,\dots,2N$, where $\ket{\bar{n}}$ and $\ket{\overline{n+N}}$ correspond to the real and imaginary parts of resonator $n$. Unbarred kets remain in the complex representation. The connection between complex-conjugate and real/imaginary representations is given by the transformations in Eq.~\eqref{eq:basis_transform},
\begin{align}
\ket*{\tilde{\psi}}&=V\ket{x},&\ket{\tilde{s}^\pm}&=V^{(e)}\ket{b^\pm},\\
\ket*{\tilde{\eta}^{(e)}}&=V^{(e)}\ket*{W^{(e)}},&\ket*{\tilde{\eta}^{(r)}}&=V^{(r)}\ket*{W^{(r)}}.\notag
\end{align}
Therefore, in the real/imaginary representation, we define
\begin{align}\label{eq:AB_operators}
A&\equiv-V^{-1}\tilde{M}_{\mathrm{eff}}V,
&A^{(s)}&\equiv-V^{-1}\tilde{M}^{(s)}V,\\
B&\equiv V^{-1}\tilde{D}V^{(e)},\qquad &B^{(e)}&\equiv V^{-1}\tilde{D}\tilde{\Theta}^{(e)}V^{(e)},\qquad B^{(r)}\equiv V^{-1}\tilde{D}^{(r)}\tilde{\Theta}^{(r)}V^{(r)},\notag
\end{align}
which are entirely real by their construction. Then, Eq.~\eqref{eq:stacked_CMT} can be written,
\begin{equation}\label{eq:RealImag_CMT}
\frac{d}{dt}\ket{x}=A\ket{x}+B\ket{b^+}+A^{(s)}\ket{x}+B^{(e)}\ket*{\dot W^{(e)}}+B^{(r)}\ket*{\dot W^{(r)}}.
\end{equation}
Since $H^{(s)}$ is linear in the jitter,
\begin{alignat}{2}\label{eq:Mj_def}
A^{(s)}&=\sum_{j=1}^{N^{(s)}}\xi^{(s)}_jA_j,&\qquad\text{with}\quad A_j&\equiv-\sigma_j^{(s)}V^{-1}\tilde{M}_jV,\\
\tilde{M}_j&\,\dot{\equiv}
\begin{pmatrix}
M_j&0\\
0&\left(M_j\right)^*
\end{pmatrix},&\qquad\text{and}\quad M_j&\equiv iH_j,\notag
\end{alignat}
where the tilde indicates the usual stacking. The third term of Eq.~\eqref{eq:RealImag_CMT} can instead be expressed as a matrix acting on the (derivative of a) vector of Wiener processes $\ket*{W^{(s)}}$,
\begin{align}
A^{(s)}\ket{x}=\sum_{j=1}^{N^{(s)}}\frac{d}{dt}\braket*{j^{(s)}}{W^{(s)}}\ket{y_j}
=\,B^{(s)}\frac{d}{dt}&\ket*{W^{(s)}},\label{eq:Wiener_proj}\\
\text{where}\quad\ket{y_j}\equiv A_j\ket{x},\qquad
B^{(s)}\equiv&\sum_{j=1}^{N^{(s)}}\ketbra*{y_j}{j^{(s)}}=\sum_{j=1}^{N^{(s)}}\sum_{\ell=1}^{2N}\ket{\bar{\ell}}\!\braket{\bar{\ell}\,}{y_j}\!\!\bra*{\,j^{(s)}}\in\mathbb{R}^{2N\times N^{(s)}}.\notag
\end{align}
Now it will be convenient to pack all Wiener processes into a single vector. To this end, introduce the filtering matrices $F^{(s)}$, $F^{(e)}$ and $F^{(r)}$, of dimensions $\mathcal{W}\times N^{(s)}$, $\mathcal{W}\times 2M$ and $\mathcal{W}\times 2N^{(r)}$ with $\mathcal{W}= N^{(s)}+2M+2N^{(r)}$, having the defining properties
\begin{align}\label{eq:filt_prop1} F^{(s)\top}F^{(s)}=\mathbb{I}_{N^{(s)}},\qquad F^{(e)\top}F^{(e)}=\mathbb{I}_{2M},\qquad F^{(r)\top}&F^{(r)}=\mathbb{I}_{2N^{(r)}},\\
F^{(s)}F^{(s)\top}+F^{(e)}F^{(e)\top}+F^{(r)}F^{(r)\top}=\mathbb{I}_\mathcal{W},\qquad F^{(a)\top}F^{(b)}&=0,\qquad a\ne b.\notag
\end{align}
The purpose of the filtering matrices is to expand $\ket{W}\equiv\sum_{a\in\{s,e,r\}} F^{(a)}\ket*{W^{(a)}}$, where the projections recover each component, $F^{(a)\top}\ket{W}=\ket*{W^{(a)}}$. For example, the full Wiener vector could be represented as $\ket{W}\doteq\begin{pmatrix}\ket*{W^{(s)}}&\ket*{W^{(e)}}&\ket*{W^{(r)}}\end{pmatrix}^\top$. Using Eq.~\eqref{eq:Wiener_proj} and the Wiener projections, Eq.~\eqref{eq:RealImag_CMT} takes the form of a Stratonovich process with drift $\mu^\circ(\ket{x},t)=A\ket{x}+B\ket{b^+}$, so the corresponding It\^o process is
\begin{align}
d\ket{x}=\mu(\ket{x},t)\,dt+\sigma(\ket{x},t)\,d\ket{W},\qquad \text{where}\quad \mu(\ket{x},t)&=\left(A+A_c\right)\ket{x}+B\ket{b^+},\label{eq:drift}\\
\sigma(\ket{x},t)&=B^{(s)}F^{(s)\top}+B^{(e)}F^{(e)\top}+B^{(r)}F^{(r)\top},\notag
\end{align}
with the noise-induced drift given by Eq.~\eqref{eq:corrected_drift},
\begin{align}\label{eq:CMT_correction_term}
A_c\ket{x}
&=\frac12\sum_{j=1}^{N^{(s)}}\Big[\underbrace{\Big(\nabla_x\,B^{(s)}F^{(s)\top}F^{(s)}\ket*{j^{(s)}}\Big)}_{\nabla_x\sigma_{:w}}\,\underbrace{\Big(B^{(s)}F^{(s)\top}F^{(s)}\ket*{j^{(s)}}\Big)}_{\sigma_{:w}}\Big]\notag\\
&=\frac12\sum_{j=1}^{N^{(s)}}\big(\nabla_x\ket{y_j}\big)\ket{y_j}
=\frac12\sum_{j=1}^{N^{(s)}}\big(\nabla_x\ket{y_j}\big)A_j\ket{x},
\end{align}
where $w$ indexes the column carrying channel $j$. The second equality uses the first property of Eq.~\eqref{eq:filt_prop1} and the definition of $B^{(s)}$ in Eq.~\eqref{eq:Wiener_proj}, and the third follows from $\ket{y_j}\equiv A_j\ket{x}$, defined in the same equation. The Jacobian $\nabla_x\ket{y_j}$ is given in the following section. The correction simplifies to $A_c\ket{x}=\tfrac12\sum_{j=1}^{N^{(s)}}A_j^2\ket{x}$ when $H_j$ is state-independent, or more generally when $A_j$ is invariant along $\ket{y_j}$, as it is for all shape perturbations treated below, and vanishes in the absence of multiplicative noise.

Physically, the correction ensures that Hermitian jitter, by itself, preserves the state norm $\braket{x}{x}$. Consider the It\^o process with only Hermitian jitter $H_j$ and $\mu^\circ=0$, so that $\mu=A_c\ket{x}$ and diffusion tensor $Q^{(s)}=\tfrac{1}{2}B^{(s)}B^{(s)\top}$. Since $H_j=H_j^\dagger$, the real $A_j$ are then antisymmetric by the definitions in Eq.~\eqref{eq:Mj_def}. Therefore, $\bra{x}A_j^\top\ket{x}=\braket{y_j}{x}=0$ identically, so differentiating yields $\bra{x}\nabla_x\ket{y_j}=-\bra{y_j}$. Then from Eq.~\eqref{eq:CMT_correction_term}, $\bra{x}A_c\ket{x}=-\tfrac{1}{2}\sum_{j=1}^{N^{(s)}}\braket{y_j}{y_j}=-\mathrm{Tr}\left[Q^{(s)}\right]$. Last, applying It\^o's lemma, Eq.~\eqref{eq:dg_w_ito}, to $\varphi=\braket{x}{x}$, since $\nabla_x\varphi=2\bra{x}$ and $\nabla_x^2\varphi=2\,\mathbb{I}_\mathcal{M}$, we have
\begin{equation}
d\!\braket{x}{x}=\left[2\bra{x}A_c\ket{x}+2\,\mathrm{Tr}\big(Q^{(s)}\big)\right]dt+2\bra{x}\sigma\ket{dW_t}=0,
\end{equation}
where the last term vanishes since, from the definition in Eq.~\eqref{eq:drift}, $\sigma\ket{dW_t}=B^{(s)}F^{(s)\top}\ket{dW_t}
=\sum_{j=1}^{N^{(s)}}\ket{y_j}\,dW_j^{(s)}$, by Eq.~\eqref{eq:Wiener_proj}, and $\braket{x}{y_j}=0$. Hence, Hermitian jitter rotates the state (pathwise) without changing its norm, whereas neglecting the correction would instead give $d\!\braket{x}{x}=2\,\mathrm{Tr}\left[Q^{(s)}\right]dt>0$, a spurious heating by the jitter alone. Equivalently, by $\braket{x}{y_j}=0$ and the chain rule of the Stratonovich convention, $d\!\braket{x}{x}=2\bra{x}\circ d\ket{x}=2\bra{x}\sigma\circ\ket{dW_t}=0$.

Since $\braket{x}{y_j}=0$, Hermitian jitter cannot resolve sensitivities to parameters that change the state norm, such as decay rates and external coupling, including through the source. These parameters act at the open channels, where additive noise is guaranteed by the fluctuation--dissipation relation and can be further instrumented at accessible ports. For the shape perturbations considered, the columns $\ket{y_j}$ lose rank at vanishing field or over a zero-measure set, e.g., phase-locked neighboring amplitudes. Near these configurations, the diffusion eigenvalue along the lost direction vanishes at least quadratically with the distance to them, so the score diverges at least inversely with that distance wherever the sensitivity is nonzero, giving heavy tails.

\subsection{Relevant Quantities for the PPGT}
Evaluating the score requires the partial derivatives of the drift and diffusion. This section gives them in a closed form for the TCMT model, along with the matrix-vector products required by the conjugate-gradient solves. In what follows we take the source to be non-trainable, though nothing prevents it, and represent the effective Hamiltonian with Kerr-type nonlinearity as
\begin{equation}
H_{\mathrm{eff}}=\sum_{n=1}^N\left[\left(\omega_n+\chi_n\vert\psi_n\vert^2\right)\ketbra{n}{n}\,+\sum_{\underset{m\ne n}{m=1}}^N\kappa_{nm}e^{i\phi_{nm}}\ketbra{n}{m}\right]-i\left[\sum_{m=1}^M\gamma_{e_m}\ketbra{n_m}{n_m}+\sum_{k=1}^{N^{(r)}}\gamma_{r_k}\ketbra{n_k}{n_k}\right],
\end{equation}
where $\omega_n$ denote resonator frequencies, $\kappa_{nm}$ and $\phi_{nm}$ denote internal coupling magnitude and phase with $\phi_{mn}=-\phi_{nm}$, $\chi_n$ denote the real-valued nonlinear Kerr coefficients, $\gamma_{r_k}$ denote radiative losses, and $\gamma_{e_m}$ denote external coupling magnitude. Correspondingly, the coupling and radiative loss matrices are
\begin{equation}\label{eq:coupmats}
D=\sum_{m=1}^{M}\sqrt{2\gamma_{e_m}}\ketbra{n_m}{e_m},\qquad D^{(r)}=\sum_{k=1}^{N^{(r)}}\sqrt{2\gamma_{r_k}}\ketbra{n_k}{r_k}.
\end{equation}
The notation, $\ket{n_m}$ ($\ket{n_k}$) and $\ket{e_m}$ ($\ket{r_k}$), used above is shorthand to indicate that lead $m$ (reservoir $k$) is coupled to resonator $n=n_m$ ($n=n_k$). When radiative loss applies to every resonator, $N^{(r)}=N$, the notation can be simplified as $k\rightarrow n$ and $\gamma_{r_k}\rightarrow\gamma_n$, and when that loss is uniform $\gamma_n\rightarrow\gamma$, as assumed in the main text. Lastly, additive noise coefficients are stored in the diagonal matrices as
\begin{equation}
\Theta^{(e)}=\sum_{m=1}^M\sigma_m^{(e)}\ketbra{e_m}{e_m},\qquad
\Theta^{(r)}=\sum_{k=1}^{N^{(r)}}\sigma_{k}^{(r)}\ketbra{r_k}{r_k}.
\end{equation}
Note that any of the noise coefficients $\sigma$ (additive or multiplicative) can be treated as training parameters, albeit with the risk of wireheading mentioned in Appendix~\ref{M-app:score} of the main text and the following caveat. Estimator variance for volatility-type parameters diverges as the time step is refined, a well-known limitation of likelihood-ratio methods with standard remedies~\cite{broadie1996estimating,glasserman2004monte}. The origin of this pathology is the diffusion terms in Eq.~\eqref{M-eq:parameter_score}, which are quadratic in $r$, and therefore $\mathcal{O}(1)$ in $\Delta t$, unlike the drift term. Integrated over a trajectory, while the estimator strictly remains unbiased, its variance accumulates as $T/\Delta t$, diverging as $\Delta t\rightarrow0$.

\subsubsection{Drift Partial Derivative}
Computing partial derivatives of the drift with respect to parameters in Eq.~\eqref{eq:drift},
\begin{align}
\partial_\theta\mu = -V^{-1}\big(\partial_\theta \tilde{M}_\mathrm{eff}\big)\ket*{\tilde{\psi}}+\left(\partial_\theta A_c\right)V^{-1}\ket*{\tilde{\psi}}+V^{-1}\big(\partial_\theta \tilde D\big)\ket{\tilde{s}^+}&=-V^{-1}\ket{\tilde{p}_\mu}+\left(\partial_\theta A_c\right)V^{-1}\ket*{\tilde{\psi}}+V^{-1}\ket{\tilde{q}_\mu},\\
\text{where}\quad\ket{p_\mu}\equiv \left(\partial_\theta M_\mathrm{eff}\right)\ket{\psi}\quad\text{and}\quad\ket{q_\mu}&\equiv\left(\partial_\theta D\right)\ket{s^+}.\notag
\end{align}
Here, the vectors $\ket{\tilde{p}_\mu}\,\dot{\equiv}\,\begin{pmatrix}\ket{p_\mu}&\ket{p_\mu}^*\end{pmatrix}^\top$ and $\ket{\tilde{q}_\mu}\,\dot{\equiv}\,\begin{pmatrix}\ket{q_\mu}&\ket{q_\mu}^*\end{pmatrix}^\top$ are stacked after multiplication, halving the compute cost since the conjugate block follows directly. The noise-induced drift sensitivity enters through $\partial_\theta A_c$, treated below.

Computing the partial derivatives explicitly by parameter type, presented with Leibniz notation for legibility, we have
\begin{align}
\frac{\partial H_\mathrm{eff}}{\partial\omega_n}&=\ketbra{n}{n},\quad&\frac{\partial H_\mathrm{eff}}{\partial \chi_n}&=|\psi_n|^2\ketbra{n}{n},\quad&\frac{\partial H_\mathrm{eff}}{\partial \kappa_{nm}}&=e^{i\phi_{nm}}\ketbra{n}{m}+e^{-i\phi_{nm}}\ketbra{m}{n},\\
\frac{\partial H_\mathrm{eff}}{\partial \gamma_{e_m}}&=-i\ketbra{n_m}{n_m},\quad&\frac{\partial H_\mathrm{eff}}{\partial \gamma_{r_k}}&=-i\ketbra{n_k}{n_k},\quad&\frac{\partial H_\mathrm{eff}}{\partial \phi_{nm}}&=i\kappa_{nm}\left[e^{i\phi_{nm}}\ketbra{n}{m}-e^{-i\phi_{nm}}\ketbra{m}{n}\right],\notag
\end{align}
and $\partial D/\partial \gamma_{e_m}=\ketbra{n_m}{e_m}/\sqrt{2\gamma_{e_m}}$ is the only non-zero $\partial_\theta D$. Notice the singularity that occurs when any $\gamma_{e_m}=0$, so the external couplings must be bounded away from zero. Additive noise parameters $\sigma^{(e)}_m$ and $\sigma^{(r)}_k$ do not enter the drift, while multiplicative noise parameters $\sigma^{(s)}_j$ enter only through the noise-induced term, whose sensitivities require the following Jacobian. Applying the product rule to the definition of $\ket{y_j}$ in Eq.~\eqref{eq:Wiener_proj}, and the chain rule through $\ket*{\tilde\psi}$,
\begin{equation}\label{eq:multiplicative_noise_jacobian}
\nabla_x\ket{y_j}=-\sigma_j^{(s)}V^{-1}\left[\tilde{M}_j+\left(\nabla_{\tilde{\psi}}\tilde{M}_j\right)\ket{\tilde\psi}\right]V,
\end{equation}
following the same Jacobian index convention introduced for $\nabla_x\sigma_{:w}$ in the Noise-Induced Drift section. The shapes for the resonator-frequency and Kerr-coefficient jitter at site $n$ are $H_j=\ketbra{n}{n}$ and $H_j=\vert\psi_n\vert^2\ketbra{n}{n}$, while the internal-coupling jitter of bond $(n,m)$ is $H_j=e^{i\phi_{nm}}\ketbra{n}{m}+e^{-i\phi_{nm}}\ketbra{m}{n}$. The second term in Eq.~\eqref{eq:multiplicative_noise_jacobian} vanishes when $M_j$ is state-independent, as it is for resonator-frequency and internal-coupling jitter. For jitter in the Kerr coefficient, it expands as
\begin{equation}
\left(\nabla_{\tilde{\psi}}\tilde{M}_j\right)\ket{\tilde\psi}=
i\lvert\psi_n\rvert^2\ketbra{n}{n}+i\psi_n^2\ketbra{n}{N+n}-i\psi_n^{*2}\ketbra{N+n}{n}-i\lvert\psi_n\rvert^2\ketbra{N+n}{N+n},
\end{equation}
and vanishes along $\tilde{M}_j\ket*{\tilde\psi}=i\vert\psi_n\vert^2\left(\psi_n\ket{n}-\psi_n^*\ket{N+n}\right)$,
\begin{equation}\label{eq:kerr_dirder_zero}
\left[\left(\nabla_{\tilde{\psi}}\tilde{M}_j\right)\ket{\tilde\psi}\right]\tilde{M}_j\ket{\tilde\psi}=
-\psi_n\lvert\psi_n\rvert^4\ket{n}+\psi_n^*\vert\psi_n\vert^2\psi_n^2\ket{n}+\vert\psi_n\vert^2\psi_n^{*2}\psi_n\ket{N+n}-\vert\psi_n\vert^4\psi_n^*\ket{N+n}=0.
\end{equation}

From Eqs.~\eqref{eq:Mj_def}, \eqref{eq:CMT_correction_term}, and \eqref{eq:multiplicative_noise_jacobian}, $A_c\ket{x}$ is quadratic in $\sigma_j^{(s)}$, so
\begin{equation}
\left(\frac{\partial A_c}{\partial\sigma_j^{(s)}}\right)\ket{x}=
\sigma_j^{(s)}V^{-1}\left[\tilde{M}_j+\big(\nabla_{\tilde{\psi}}\tilde{M}_j\big)\ket*{\tilde\psi}\right]\tilde{M}_j\ket*{\tilde\psi}
\end{equation}
in general. By Eq.~\eqref{eq:kerr_dirder_zero} for Kerr-coefficient jitter, and the rest state-independent, it simplifies to $\big[\partial A_c/\partial \sigma_j^{(s)}\big]\ket{x}=
\sigma_j^{(s)}V^{-1}\tilde{M}_j^2\ket*{\tilde\psi}$ for each shape perturbation considered here.

For other parameters that enter $H_j$,
\begin{equation}
\left(\partial_\theta A_c\right)\ket{x}=\frac12\sum_{j=1}^{N^{(s)}}\left(\sigma_j^{(s)}\right)^2V^{-1}\left\{\left[\partial_\theta\tilde{M}_j+\big(\nabla_{\tilde{\psi}}\,\partial_\theta\tilde{M}_j\big)\!\ket*{\tilde\psi}\right]\tilde{M}_j+\left[\tilde{M}_j+\big(\nabla_{\tilde{\psi}}\tilde{M}_j\big)\ket*{\tilde\psi}\right]\partial_\theta\tilde{M}_j\right\}\ket*{\tilde\psi},
\end{equation}
in general, but vanishes for each shape perturbation considered here, since the phase in the internal-coupling jitter cancels in $H_j^2=\ketbra{n}{n}+\ketbra{m}{m}$ and the others are parameter independent.

Similarly, we compute the partial derivatives of the drift with respect to state in Eq.~\eqref{eq:drift}. Since $B$ is a linear operator,
\begin{equation}
\nabla_x \mu=\left(A+A_c\right)+\left[\nabla_x \left(A+A_c\right)\right]\ket{x}.
\end{equation}
The second term is non-zero only for state-dependent operators. The first part is the nonlinear contribution from the physical drift,
\begin{equation}
\left(\nabla_x A\right)\ket{x}=-V^{-1}\left[\left(\nabla_{\tilde{\psi}}\tilde{M}_\mathrm{eff}\right)\ket{\tilde{\psi}}\right]V.
\end{equation}
For Kerr-type nonlinearity, the bracketed operator expands as
\begin{equation}
\left(\nabla_{\tilde{\psi}} \tilde{M}_\mathrm{eff}\right)\ket{\tilde{\psi}}=i\sum_n\left[\chi_n\lvert\psi_n\rvert^2\ketbra{n}{n}+\chi_n\psi_n^2\ketbra{n}{N+n}-\chi_n^*\psi_n^{*2}\ketbra{N+n}{n}-\chi_n^*\lvert\psi_n\rvert^2\ketbra{N+n}{N+n}\right],
\end{equation}
with conjugates kept for generality, though the Kerr coefficients are taken real.

The second part, from the noise-induced drift, combines with $A_c$ by the product rule applied to Eq.~\eqref{eq:CMT_correction_term},
\begin{equation}
A_c+\left(\nabla_x A_c\right)\ket{x}=\nabla_x\left(A_c\ket{x}\right)=\frac12\sum_{j=1}^{N^{(s)}}\left[\left(\nabla_x^2\ket{y_j}\right)\ket{y_j}+\left(\nabla_x\ket{y_j}\right)^2\right],
\end{equation}
where the Hessian is contracted with $\ket{y_j}$ along one of its symmetric derivative indices. For Kerr-coefficient jitter at site $n$, carried by multiplicative noise channel $j$, the contracted Hessian expands as
\begin{equation}
\left(\nabla_x^2\ket{y_j}\right)\ket{y_j}=-2\left(\sigma_j^{(s)}\right)^2\lvert\psi_n\rvert^2\,V^{-1}\left(\psi_n^2\ketbra{n}{N+n}+\psi_n^{*2}\ketbra{N+n}{n}\right)V.
\end{equation}
For state-independent $H_j$ the Hessian vanishes and $\nabla_x\ket{y_j}=A_j$, recovering the simplified form of $A_c$ beneath Eq.~\eqref{eq:CMT_correction_term}.

\subsubsection{Diffusion Multiplication}
The conjugate gradient solves introduced in the Inverse-Free Score Evaluation section require the matrix-vector product $Q\ket{v}$, where $\ket{v}\equiv\sum_\ell v_\ell\ket{\overline{\ell}}$ is arbitrary. Here, we construct it in a closed form, avoiding materialization of $Q$. From Eq.~\eqref{eq:filt_prop1}, the diffusion tensor can be expressed as
\begin{equation}\label{eq:Diffusion}
Q\equiv\frac{\sigma\sigma^\top}{2}=\frac{1}{2}\left[B^{(s)}B^{(s)\top}+B^{(e)}B^{(e)\top}+B^{(r)}B^{(r)\top}\right].
\end{equation}
Focusing on the multiplicative noise terms first, from Eq.~\eqref{eq:Wiener_proj},
\begin{align}\label{eq:mult_noise_terms}
B^{(s)}B^{(s)\top}&=\sum_{j=1}^{N^{(s)}}\ketbra{y_j}{y_j}=\sum_{j=1}^{N^{(s)}}\left(\sigma_j^{(s)}\right)^2 V^{-1}\ketbra{\tilde{p}_j}{\tilde{p}_j}V^{\dagger-1},\\
\text{where}\quad\ket{p_j}&\equiv M_j\ket{\psi}\notag\\
\text{and}\quad\ket{\tilde{p}_j}&\,\dot{\equiv}\,\begin{pmatrix}
\ket{p_j}&
\ket{p_j}^*
\end{pmatrix}^\top.\notag
\end{align}
Accordingly, the column entering Eq.~\eqref{eq:trace_term} for multiplicative noise channel $j$ is $\sigma_{:w}=\ket{y_j}=-\sigma_j^{(s)}V^{-1}\ket{\tilde {p}_j}$.

Therefore,
\begin{equation}
B^{(s)}B^{(s)\top}\ket{v}=\sum_{\ell=1}^{2N}\sum_{j=1}^{N^{(s)}}v_\ell\left(\sigma_j^{(s)}\right)^2 V^{-1}\ketbra{\tilde{p}_j}{\tilde{p}_j}V^{\dagger-1}\ket{\overline{\ell}}=\sum_{\ell=1}^{2N}\sum_{j=1}^{N^{(s)}}v_\ell\left(\sigma_j^{(s)}\right)^2 \bra{\overline{\ell}}\,V^{-1}\ket{\tilde{p}_j}V^{-1}\ket{\tilde{p}_j},
\end{equation}
where in the last step, we used the fact that the scalar inner product is real.

Next, turn to the additive noise terms, for which we will use the following blockwise identities,
\begin{equation}\label{eq:transform_identities}
\begin{pmatrix}
\mathbb{I}&i\mathbb{I}\\
\mathbb{I}&-i\mathbb{I}
\end{pmatrix}^{-1}
=\frac{1}{2}\begin{pmatrix}
\mathbb{I}&\mathbb{I}\\
-i\mathbb{I}&i\mathbb{I}
\end{pmatrix},
\qquad
\begin{pmatrix}
\mathbb{I}&i\mathbb{I}\\
\mathbb{I}&-i\mathbb{I}
\end{pmatrix}
\begin{pmatrix}
\mathbb{I}&i\mathbb{I}\\
\mathbb{I}&-i\mathbb{I}
\end{pmatrix}^\top
=\begin{pmatrix}
\mathbb{I}&i\mathbb{I}\\
\mathbb{I}&-i\mathbb{I}
\end{pmatrix}
\begin{pmatrix}
\mathbb{I}&\mathbb{I}\\
i\mathbb{I}&-i\mathbb{I}
\end{pmatrix}
=2\begin{pmatrix}
0&\mathbb{I}\\
\mathbb{I}&0
\end{pmatrix},
\end{equation}
where $\mathbb{I}$ carries the dimension of the corresponding transform in Eq.~\eqref{eq:basis_transform}. From Eqs.~\eqref{eq:AB_operators} and \eqref{eq:transform_identities}, we have
\begin{align}
B^{(e)}B^{(e)\top}\ket{v}&=\sum_{\ell=1}^{2N}v_\ell\,V^{-1}CV^{-1\top}\ket{\overline{\ell}},\\
\text{where}\quad C&\equiv\tilde{D}\tilde\Theta^{(e)}V^{(e)}V^{(e)\top}\tilde\Theta^{(e)}\tilde{D}^\top\doteq
2\begin{pmatrix}
0&D(\Theta^{(e)})^2D^\top\\
D(\Theta^{(e)})^2D^\top&0
\end{pmatrix}.\notag
\end{align}
Furthermore, performing the multiplication explicitly from Eqs.~\eqref{eq:coupmats} and \eqref{eq:transform_identities},
\begin{align}
V^{-1}CV^{-1\top}&\doteq
\begin{pmatrix}
D\left(\Theta^{(e)}\right)^2D^\top&0\\
0&D\left(\Theta^{(e)}\right)^2D^\top
\end{pmatrix},\\
\text{where}\quad D\left(\Theta^{(e)}\right)^2D^\top&=\sum_{m=1}^M 2\gamma_{e_m}\left(\sigma_m^{(e)}\right)^2\ketbra{n_m}{n_m}.\notag
\end{align}
Hence, for the external coupling and similarly for the radiative terms,
\begin{align}
B^{(e)}B^{(e)\top}\ket{v}&=\sum_{m=1}^M2\gamma_{e_m}\left(\sigma_m^{(e)}\right)^2\left(v_{n_m}\,\ket{\overline{n_m}}+v_{n_m+N}\,\ket{\overline{n_m+N}}\right),\notag\\ B^{(r)}B^{(r)\top}\ket{v}&=\sum_{k=1}^{N^{(r)}}2\gamma_{r_k}\left(\sigma_k^{(r)}\right)^2\left(v_{n_k}\,\ket{\overline{n_k}}+v_{n_k+N}\,\ket{\overline{n_k+N}}\right),
\end{align}
with the corresponding columns entering Eq.~\eqref{eq:trace_term},
$\sigma_{:w}=\sqrt{2\gamma_{e_m}}\,\sigma_m^{(e)}\ket{\overline{n_m+\mathfrak{q}N}}$ and $\sqrt{2\gamma_{r_k}}\,\sigma_k^{(r)}\ket{\overline{n_k+\mathfrak{q}N}}$ for lead $m$ and reservoir $k$, respectively, where $\mathfrak{q}\in\{0,1\}$ selects the real or imaginary quadrature.

\subsubsection{Diffusion Partial Derivative}
From the results of the previous section, we immediately have the associated multiplicative partial derivative,
\begin{equation}
\frac{\partial Q}{\partial \sigma_j^{(s)}}\ket{v}=\sum_{\ell=1}^{2N}v_\ell\sigma_j^{(s)} \bra{\overline{\ell}}\,V^{-1}\ket{\tilde{p}_j}V^{-1}\ket{\tilde{p}_j}.
\end{equation}
For jitter in the internal coupling of bond $(n,m)$, carried by multiplicative noise channel $j$, the shape perturbation depends on the trainable phase $\phi_{nm}$, so that
\begin{align}
\frac{\partial Q}{\partial\phi_{nm}}\ket{v}&=\frac{1}{2}\left(\sigma_j^{(s)}\right)^2\sum_{\ell=1}^{2N}v_\ell \left[\bra{\overline{\ell}}\,V^{-1}\ket{\partial_\theta \tilde{p}_j}V^{-1}\ket{\tilde{p}_j}+\bra{\overline{\ell}}\,V^{-1}\ket{\tilde{p}_j}V^{-1}\ket{\partial_\theta\tilde{p}_j}\right],\\
\text{where}\quad\ket{\partial_\theta{p}_j}&=i(\partial_\theta H_j)\ket{\psi}=-\left(e^{i\phi_{nm}}\psi_m\ket{n}-e^{-i\phi_{nm}}\psi_n\ket{m}\right)\quad\text{for}\quad\theta=\phi_{nm},\notag\\
\text{and}\quad\ket{\partial_\theta\tilde{p}_j}&\,\dot{\equiv}\,\begin{pmatrix}
\ket{\partial_\theta p_j}&
\ket{\partial_\theta p_j}^*
\end{pmatrix}^\top.\notag
\end{align}
The corresponding column partials entering Eq.~\eqref{eq:trace_term} are $\partial_\theta\sigma_{:w}=-V^{-1}\ket{\tilde{p}_j}$, for $\theta=\sigma_j^{(s)}$, $\partial_\theta\sigma_{:w}=-\sigma_j^{(s)}V^{-1}\ket{\partial_\theta\tilde p_j}$, for $\theta=\phi_{nm}$ of the jittered bond, and otherwise zero for the multiplicative noise channels.

As for the additive terms, we have
\begin{align}
\frac{\partial Q}{\partial \sigma_m^{(e)}}\ket{v}&=2\gamma_{e_m}\sigma_m^{(e)}\left(v_{n_m}\ket{\overline{n_m}}+v_{n_m+N}\ket{\overline{n_m+N}}\right),&\frac{\partial Q}{\partial \gamma_{e_m}}\ket{v}&=\left(\sigma_m^{(e)}\right)^2\left(v_{n_m}\ket{\overline{n_m}}+v_{n_m+N}\ket{\overline{n_m+N}}\right),\notag\\
\frac{\partial Q}{\partial \sigma_k^{(r)}}\ket{v}&=2\gamma_{r_k}\sigma_k^{(r)}\left(v_{n_k}\ket{\overline{n_k}}+v_{n_k+N}\ket{\overline{n_k+N}}\right),&\frac{\partial Q}{\partial \gamma_{r_k}}\ket{v}&=\left(\sigma_k^{(r)}\right)^2\left(v_{n_k}\ket{\overline{n_k}}+v_{n_k+N}\ket{\overline{n_k+N}}\right).
\end{align}
The corresponding column partials entering Eq.~\eqref{eq:trace_term} are
\begin{align}
\frac{\partial \sigma_{:w}}{\partial \sigma_m^{(e)}}&=\sqrt{2\gamma_{e_m}}\,\ket{\overline{n_m+\mathfrak{q}N}},&\frac{\partial \sigma_{:w}}{\partial \gamma_{e_m}}&=\frac{\sigma_m^{(e)}}{\sqrt{2\gamma_{e_m}}}\ket{\overline{n_m+\mathfrak{q}N}},\notag\\
\frac{\partial \sigma_{:w}}{\partial \sigma_k^{(r)}}&=\sqrt{2\gamma_{r_k}}\,\ket{\overline{n_k+\mathfrak{q}N}},&\frac{\partial \sigma_{:w}}{\partial \gamma_{r_k}}&=\frac{\sigma_k^{(r)}}{\sqrt{2\gamma_{r_k}}}\ket{\overline{n_k+\mathfrak{q}N}},
\end{align}
where, as before, $\mathfrak{q}\in\{0,1\}$ selects the real or imaginary quadrature.

Likewise, we now construct the partial derivative of the diffusion with respect to state. The additive noise terms are zero, leaving only the multiplicative terms non-zero, so from Eqs.~\eqref{eq:Diffusion} and \eqref{eq:mult_noise_terms},
\begin{equation}
\nabla_xQ=\frac{1}{2}\sum_{j=1}^{N^{(s)}}\left[\nabla_x\ket{y_j}\otimes\bra{y_j}+(\nabla_x\ket{y_j}\otimes\bra{y_j})^\top\right],
\end{equation}
where $\otimes$ denotes the outer product, appending the state index carried by $\nabla_x$, the transpose is with respect to the two matrix indices of $Q$, and $\nabla_x\ket{y_j}$ is given in Eq.~\eqref{eq:multiplicative_noise_jacobian}.

\newpage

\section{Optimization Details}

\subsection{Trajectory Budget Escalation Details}

\begin{figure}[H]
\centering
\includegraphics[width=\columnwidth]{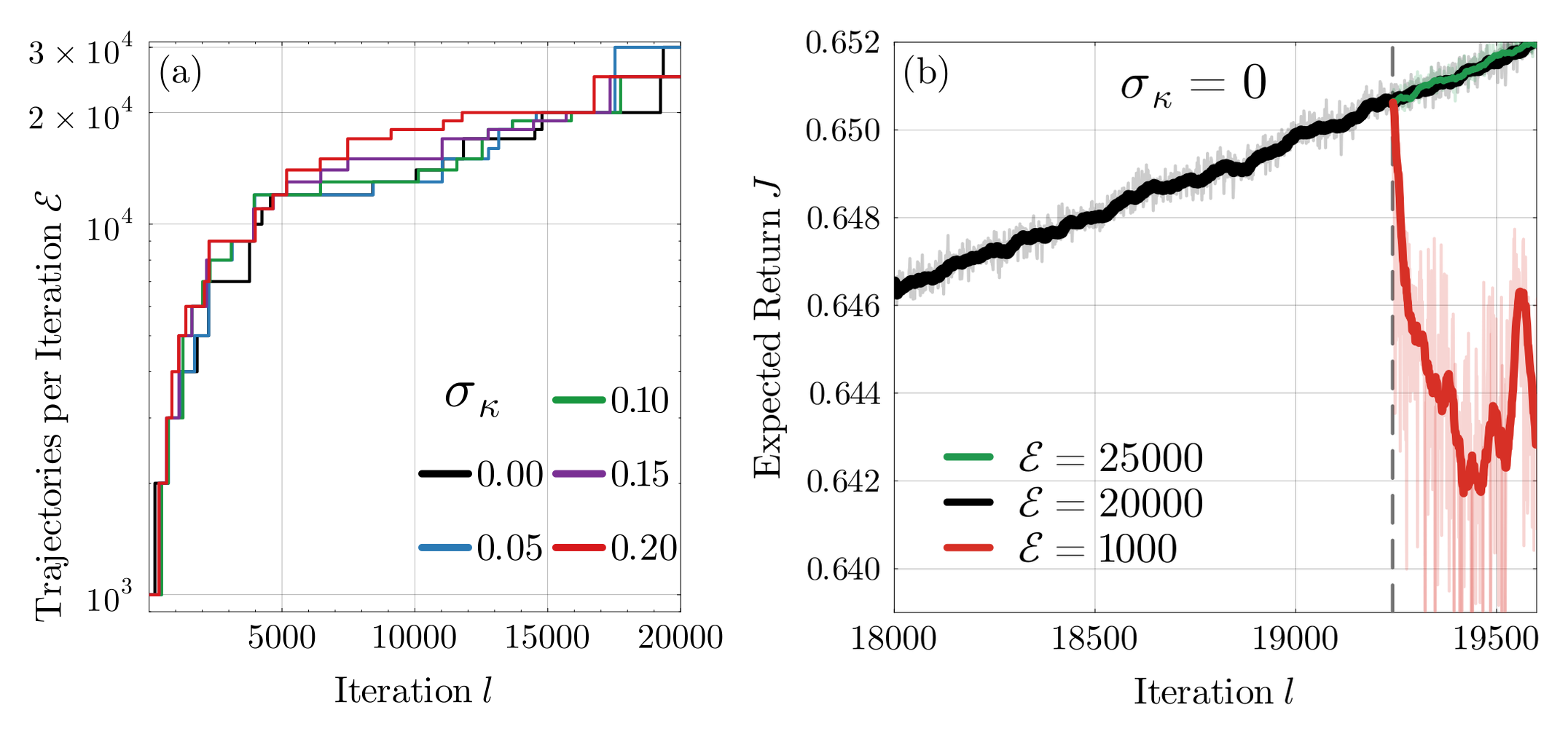}
\caption{\label{fig:budget}
  \textbf{Trajectory Budget}
  \textbf{(a)}~Trajectories per iteration versus training iteration for $\sigma_\kappa$ values indicated by the legend. \textbf{(b)}~Continuations of the $\sigma_\kappa=0$ optimization from a common breakpoint (vertical dashed line) at escalated (green), constant (black), and de-escalated (red, which departs from the gradient flow) trajectory budgets (legend), overlaid with $25$-iteration rolling averages.}
\end{figure}

The trajectory budget per iteration is plotted in Fig.~\ref{fig:budget}(a), while Fig.~\ref{fig:budget}(b) shows the role of ensemble size escalation: a fixed trajectory budget caps the achievable optimum by its noise floor, so some escalation is required to sustain progress, although later escalations were over-provisioned.

Escalation triggered when the trend over the trailing $\Delta l_{\mathrm{conv}} = 100$ iterations became statistically indistinguishable from zero at the $95\%$ confidence level, intended to indicate that progress had become saturated by the noise floor. The optimizer then rewound to the onset of the plateau and resumed at the next ensemble size. The onset was determined by extrapolating a quadratic fit to the recent objective history ($[\,l - 12\Delta l_{\mathrm{conv}},\ l - 2\Delta l_{\mathrm{conv}}\,]$) to its rising intersection with the plateau mean, falling back to $l - \Delta l_{\mathrm{conv}}$ when no admissible breakpoint exists, re-running the tier from its start when the breakpoint lands within $\Delta l_\mathrm{conv}$ of its own onset, or escalating in place before the fit window exists. In practice, the escalation parameters were set too sensitively, which led to spurious over-escalation: at the set learning rate, progress within $\Delta l_\mathrm{conv}$ iterations can be slow and stochastic enough that the trend test cannot reliably detect it. The fit window multiples were likewise chosen ad hoc, and both could benefit from tuning against the learning rate.

\subsection{Adaptive Conditioning for Heavy-Tailed Intervals}

State-dependent baselines for enhanced variance reduction were conditioned by partitioning the support of the total intensity at sites that interact with the corresponding parameter at time $t$ into $K$ equipopulation intervals, $\left(q_{k-1},q_k\right]$ with $q_{k-1}<q_k$, $k=1,\dots,K$, across the ensemble. For example, baseline conditioning associated with $\kappa_{nm}$ or $\phi_{nm}$ depends on which interval contains $\lvert\psi_n(t)\rvert^2+\lvert\psi_m(t)\rvert^2$, $\beta(x)=k\Leftrightarrow q_{k-1}<x\leq q_k$. For on-site parameters, the partitioning depends only on the intensities at that site, $\lvert\psi_n\left(t\right)\rvert^2$. More generally, the state-dependent baseline takes the form,
\begin{equation}
\bar\lambda\left(x,t\right)=\frac{\mathbb{E}\left[\hat\lambda\,U^2\Big\vert t, \beta\!\left(x\right)\right]}{\mathbb{E}\left[U^2\big\vert t, \beta\!\left(x\right)\right]},
\end{equation}
where $U$ denotes the corresponding score (leave-one-out trajectory index notation suppressed). The baseline is estimated from a finite ensemble, and therefore is itself noisy. Finer partitions leave fewer samples per interval, and we observe this can give sub-optimal variance reduction compared with a coarser partition, especially under multiplicative noise and strong nonlinearity. To compensate, coarser conditioning can pool these intervals into larger equipopulation sets $\mathcal{B}$ at the cost of conditioning resolution. The granularity can be cheaply selected on-the-fly, up to some pre-selected maximum resolution, by collapsing the finely conditioned expectations using the tower property,
\begin{equation}
\frac{\mathbb{E}\left[\hat\lambda\,U^2\Big\vert t,\mathcal{B}\right]}{\mathbb{E}\left[U^2\big\vert t,\mathcal{B}\right]}=
\frac{\sum_{\beta\in \mathcal{B}}\mathbb{E}\left[\hat\lambda\,U^2\Big\vert t,\beta\right]}{\sum_{\beta\in \mathcal{B}}\mathbb{E}\left[U^2\Big\vert t,\beta\right]},
\end{equation}
which follows simply in the case of equipopulation intervals, where the interval probability is constant, $P\left(\beta(x)\right)=1/K$. Results in the main text used additive Wiener processes, and the selection saturated at the maximum interval resolution with $K=20$.

\bibliography{bibtex}